\documentclass[10pt]{article}
\usepackage{mhchem}
\usepackage{fancyhdr}
\usepackage{extramarks} 
\usepackage{amsmath}
\usepackage{amsthm}
\usepackage{amsfonts}
\usepackage{siunitx}
\usepackage{tikz}
\usepackage[plain]{algorithm}
\usepackage{algpseudocode}
\usepackage{multirow}
\usepackage{booktabs}
\usepackage{palatino}
\usepackage{graphicx}
\usepackage{subfigure}
\usepackage[colorlinks,linkcolor=black,anchorcolor=black,citecolor=black,urlcolor=blue]{hyperref}
\usepackage{amsmath,bm}
\usepackage{booktabs}
\usepackage{mathtools}
\usepackage{amssymb}
\usepackage{caption}
\usepackage{capt-of}
\usepackage{mciteplus}
\usepackage{cite}
\usepackage{mathrsfs}
\usepackage[title,titletoc,toc]{appendix}
\usepackage{xr}
\usepackage{parskip}
\usepackage{soul}
\usepackage{textcomp}
\usepackage[colaction]{multicol}
\usepackage[switch]{lineno}
\usepackage{lipsum}
\usepackage{etoolbox}
\usepackage{longtable}
\usepackage{array}
\usepackage{tablefootnote}
\usepackage{graphicx}
\usepackage{color, colortbl}
\usepackage{hhline}

\definecolor{Lightblue}{rgb}{0.867,0.914,0.961}
\definecolor{Lightgreen}{rgb}{0.883,0.934,0.848}

\newcolumntype{C}[1]{>{\centering\arraybackslash}p{#1}}
\usetikzlibrary{automata,positioning}
\usetikzlibrary{automata,positioning}

\begin{document}

\title{Proteome-informed machine learning studies of cocaine addiction}

 \author{ Kaifu Gao$^1$, Dong Chen$^1$, Alfred J  Robison$^2$, and Guo-Wei Wei$^{1,3,4}$\footnote{
 		Corresponding author.		Email: weig@msu.edu} \\
 $^1$ Department of Mathematics, \\
 Michigan State University, MI 48824, USA.\\
 $^2$ Department of Physiology,\\
Michigan State University, MI 48824, USA.\\
$^3$Department of Electrical and Computer Engineering,\\
 Michigan State University, MI 48824, USA. \\
 $^4$ Department of Biochemistry and Molecular Biology,\\
 Michigan State University, MI 48824, USA. \\
 }

\date{\today} 

\maketitle

\begin{abstract}
Cocaine addiction accounts for a large portion of substance use disorders and threatens millions of lives worldwide. There is an urgent need to come up with efficient anti-cocaine addiction drugs. Unfortunately, no medications have been approved by the Food and Drug Administration (FDA), despite the extensive effort in the past few decades. The main challenge is the intricate molecular mechanisms of cocaine addiction, involving synergistic interactions among proteins upstream and downstream of dopamine transporter (DAT) functions impacted by cocaine. However, traditional \textit{in vivo} or \textit{in vitro} experiments can not address the roles of so many proteins, highlighting the need for innovative strategies in the field. We propose a proteome-informed machine learning/deep learning (ML/DL) platform to discover nearly optimal anti-cocaine addiction lead compounds. We construct and analyze proteomic protein-protein interaction (PPI) networks for cocaine dependence to identify 141 involved drug targets and represent over 60,000 associated drug candidates or experimental drugs in the latent space using an autoencoder (EA) model trained from over 104 million molecules. We build 32 ML models for cross-target analysis of these drug candidates for side effects and repurposing potential. We further screen the absorption, distribution, metabolism, excretion, and toxicity (ADMET) properties of these candidates. Our platform reveals that essentially all of the existing drug candidates, including dozens of experimental drugs, fail to pass our cross-target and ADMET screenings. Nonetheless, we have identified two nearly optimal leads for further optimization. 
 

\end{abstract}


\pagenumbering{roman}
\begin{verbatim}
\end{verbatim}

%
\newpage

\setcounter{page}{1}
\renewcommand{\thepage}{{\arabic{page}}}

\section{Introduction}


Substance use disorders (SUD) involving alcohol, opioids, cocaine, etc., adversely affect a growing population of individuals and families worldwide, constituting a significant socioeconomic burden with increasing medical expenses and crime. Psychostimulants, especially cocaine, account for a large portion of SUD and impact millions of lives. In the  United States (US) alone, among the 70,630 SUD-related deaths in 2019, 15,883 were due to cocaine addiction. The hazard from cocaine addiction and subsequent mortality calls for effective medications. However, currently no cocaine addiction medications have been approved by the US Food and Drug Administration (FDA) \cite{newman2021new}.


The psychotropic properties of cocaine primarily derive from blocking the dopamine transporter (DAT). Specifically, cocaine blocks DAT and prevents dopamine reuptake from the synaptic cleft into the pre-synaptic axon terminal. As a result, a higher dopamine level in the synaptic cleft promotes the activation of dopamine receptors in the post-synaptic neuron, which generates euphoria and arousal \cite{cheng2015insights}. Among these dopamine receptors, the D$_3$ dopamine receptor (D$_3$R) plays a critical role in the reward and addiction of cocaine since D$_3$R has a large population in the mesolimbic reward system \cite{matuskey2014dopamine}. Therefore, D$_3$R may be  an important target for treating drug addiction. Among other dopamine receptors, D$_1$R and D$_2$R are the most abundant ones in the brain. D$_1$R along with other D$_1$-like receptors stimulates intracellular cyclic adenosine monophosphate (cAMP) levels \cite{baik2013dopamine}. The functions of D$_1$-like receptors are to regulate the growth of neurons, some D$_2$R-mediated events, as well as other behaviors \cite{paul1992d1}.  D$_2$R, D$_3$R, and D$_4$R belong to D$_2$-like receptors and inhibit intracellular cAMP levels. D$_2$R intimately joins in the circuitry of motor control, and it is the main target of most antipsychotic drugs \cite{RAMANATHAN2018}. D$_4$R relates to many neurological and psychiatric conditions \cite{ptavcek2011dopamine} including schizophrenia and bipolar disorder, attention-deficit/hyperactivity disorder (ADHD), addictive behaviors, Parkinson's disease, and eating disorders such as anorexia nervosa. Some studies also suggest that D$_1$R, D$_2$R, and D$_4$R are involved in  locomotor activity changes induced by cocaine and other psychostimulants  \cite{di2014dopamine}.

Cocaine also blocks the serotonin transporter and norepinephrine transporter, inhibiting the reuptake of serotonin and norepinephrine and thus increasing activation of serotonin and norepinephrine receptors. Repeated use of psychostimulants alters gene expression throughout the brain, including in the nucleus accumbens, a critical center for reward processing. Frequent cocaine exposure increases expression of the transcription factor  $\Delta$FosB and brain-derived neurotrophic factor (BDNF), which in turn regulate gene expression to alter both dendritic and synaptic morphology and function in the nucleus accumbens and prefrontal cortex \cite{hope1998cocaine}, likely driving the long-term compulsion for drug seeking and taking that underlies addiction \cite{robison2011transcriptional}.

 
Currently, experimental medications against cocaine addiction mainly target DAT and D$_3$R \cite{newman2021new}. (1) Atypical DAT inhibitors are studied widely. While cocaine and its analogs (typical DAT inhibitors) bind and stabilize outward-facing conformations of DAT \cite{schmitt2013nonclassical}, atypical DAT inhibitors stabilize inward-facing conformations of DAT upon their binding. DAT with an inward-facing DAT conformation is much harder for cocaine to bind (an approximate 100-fold loss in the potency of cocaine for the inward-facing conformation comparing with the outward-facing conformation) \cite{reith2015behavioral,schmitt2013nonclassical}. In other words, even binding affinities (BAs) to DAT are weaker than that of cocaine, the pretreatment by atypical inhibitors can still prevent DAT from being blocked by cocaine. More importantly, atypical DAT inhibitors do not induce cocaine-like behaviors or addiction \cite{reith2015behavioral}. (2) Another promising approach against cocaine addiction is D$_3$R antagonists/partial agonists. D$_3$R antagonists could effectively attenuate the motivation to earn psychostimulants and reduce relapse-related behaviors.  D$_3$R partial agonists can not only functionally block the effect of cocaine addiction but also elicit the partial activation of their receptor targets under abstinence conditions, and thus potentially mitigate withdrawal effects \cite{jordan2019progress}. 

Besides potency, the safety of cocaine addiction treatments must be carefully evaluated. One dangerous side-effect target for drug addiction treatments is the human \textit{ether-a-go-go} (hERG) potassium channel, which could incur adverse side effects and even death. hERG generates the delayed rectifying potassium current. When a compound inhibits the hERG channel, it interferes with potassium current, prolongs QT interval, and results in torsades de pointes (TdP), a potentially lethal ventricular tachycardia  \cite{hancox2008herg}. Thus, hERG poses a serious challenge to drug development since it can easily attract small compounds, especially those with protonatable amines and aromatic groups, a hallmark of many neurotransmitter transport inhibitors and GPCR ligands \cite{cavalluzzi2020human}. The hERG blockade was a popular reason for drug withdrawals in the 1990s and early 2000s. Therefore, in early 2000, FDA included hERG side effect in their updated regulations: the TdP liability of drug candidates have to be evaluated in\textit{-vivo} or {\it in} \textit{vitro} in phase 1 clinical trials \cite{food2005international}.

\begin{figure}[ht!]
	\centering
	\includegraphics[width=0.9\textwidth]{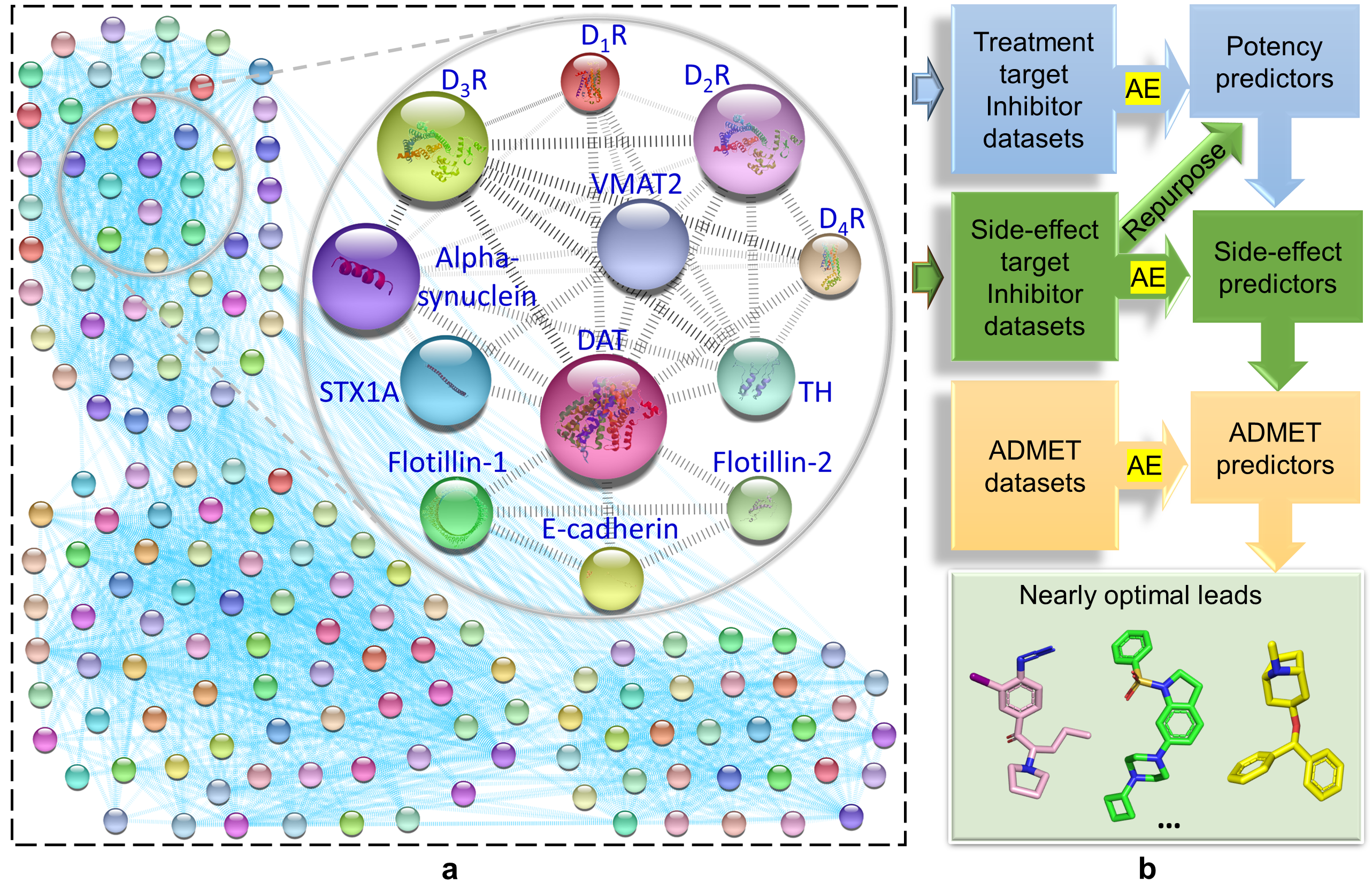}
	\caption{ DAT-centered global and core PPI networks and proteome-informed ML workflow for anti-cocaine addiction drug discovery. The datasets inferred   by the PPI networks are collected, represented by latent-vector fingerprint (LV-FP) via an autoencoder (AE), and used to construct BA predictors. The hits obtained from potency predictors, including those repurposing hits, are screened for potential side effects. The resulting promising candidates are further evaluated for 
	ADMET properties to discover nearly optimal anti-cocaine addiction  leads.
 Abbreviations: dopamine transporter (DAT), dopamine receptor D1 (D$_1$R), dopamine receptor D2 (D$_2$R), dopamine receptor D3 (D$_3$R), dopamine receptor D4 (D$_4$R), syntaxin-1A (STX1A), tyrosine hydroxylase (TH), and vesicular monoamine transporter 2 (VMAT2). }
	\label{fig:workflow}
\end{figure}

The mechanism of cocaine addiction is very complicated, involving far more targets than DAT, $D_3$R, and hERG. All the proteins upstream and downstream of DAT functions could be impacted by cocaine, which covers a large number of proteins and interactions as shown in Figure \ref{fig:workflow}a. On the one hand, these proteins can become potential treatment targets for cocaine addiction. On the other hand, blocking these proteins also probably brings cocaine-like symptoms or other severe off-target effects. Therefore, these proteins could be critical sources of side effects. Thus, we need to systematically  investigate potential compounds inhibiting different cocaine-addiction targets, as well as the putative side effects from agents blocking these targets.


One method to systematically unveil potential treatment and critical side-effect targets is to examine sizable protein-protein interaction (PPI) networks on the proteome scale. A PPI network accounts for not only direct (physical/chemical) interactions but also indirect (functional) association \cite{szklarczyk2019string}, in which a connection represents two proteins jointly contributing to a specific biological function even without direct physical/chemical interaction. As a result, a proteomic PPI network is a suitable tool for systematically searching a large number of proteins relating to a specific disease, providing a ``pool'' of potential treatments and critical side-effect targets, such as cocaine addiction in the present work.   The String v11 database collects a large number of protein-protein interaction involving a total of 24,584,628 proteins from 5,090 organisms.\cite{szklarczyk2019string}. One can extract the large-scale PPI network for a specific human protein from String. 

Traditional \textit{in vivo} or \textit{in vitro} are too time-consuming and expensive to test all the proteins in a proteomic PPI network efficiently. Additionally, large-scale experiments on animals could raise important ethical concerns. In the circumstance of large-scale assays, machine learning/deep learning (ML/DL) technologies are promising, at least for initial evaluation and screening. ML/DLs have been widely applied at different stages of drug design and discovery  \cite{ghislat2021recent}. ML/DL could help to predict drug potency \cite{ghislat2021recent}, repurpose existing drugs to new diseases \cite{gao2020repositioning}, and even generate new drug-like compounds for further screening \cite{gao2020generative}. ML/DL methods have also been established for  lead optimization of various druggable properties \cite{daina2017swissadme,gao20202d}, including solubility, partition coefficient, toxicity, pharmacokinetics, and pharmacodynamics. These technologies could largely reduce the need for time-consuming and expensive experiments and thus speed up drug discovery, significantly benefiting human health and welfare.


In this work, we designed a proteome-informed ML/DL workflow to discover nearly optimal anti-cocaine addiction leads, as shown in Figure \ref{fig:workflow}. First, we extracted a proteomic PPI network of cocaine addiction from the String database \cite{szklarczyk2019string} to infer 141 potential treatment and critical side-effect targets. Although more side-effect targets outside the PPI network should be considered, we limit our effort to critical ones revealed by the PPI network in this work. Second, for targets in the network, the associated molecules are presented by latent-vector fingerprints (LV-FPs) via an autoencoder (AE) and  
built ML/DL-based BA predictors. Third,  we carry out cross-target BA predictions of over 60,000 associated compounds to screen possible side effects and repurposing potentials. Interestingly, the correlation between predicted BAs for different targets could reveal binding-site similarities of among targets, which is a byproduct of our proteome-informed ML/DL workflow. Finally, we applied ML-based models to further evaluate the pharmacokinetic properties, i.e.,  absorption, distribution, metabolism,  excretion, and  toxicity (ADMET) as well as synthesizability. These evaluations, together with the potency and side-effect analysis, form a series of  filters to screen nearly optimal lead compounds. 




\section{Results}

\subsection{The cocaine addiction PPI networks } \label{section:ppi-network}

 
DAT is well known as the critical direct target of cocaine. To study the PPI network of cocaine addiction, we input ``DAT'' to the String database  and extracted a global network and a core network of DAT interactions (see Figure \ref{fig:workflow}a). The global network contains 141 nodes and 1696 edges. The core network only considers the proteins having direct known interactions with DAT, which leads to 11 nodes and 29 edges.  The global network could be  decomposed into three clusters, implying these proteins involve three different primary functions. The core network resides in one cluster, with 11 critical proteins in the biochemical pathways of cocaine addiction.


Apart from DAT, VMAT2 is another critical node in the core network. VMAT2 is a transport protein integrated into the membrane of synaptic vesicles of presynaptic neurons. Its main function is to transport monoamines, especially neurotransmitters such as dopamine, norepinephrine, serotonin, and histamine, from the cytosol into synaptic vesicles, which then release the neurotransmitters into synapses as chemical messages to postsynaptic neurons. Many psychostimulants such as cocaine interact with VMAT2, which emphasizes its clinical significance  \cite{brown2001regulation}. Moreover, alpha-synuclein is a neuronal protein that plays several roles in synaptic activity, such as regulation of synaptic vesicle trafficking and subsequent neurotransmitter release. It participates as a monomer in synaptic vesicle exocytosis by enhancing vesicle priming, fusion, and dilation of exocytotic fusion pores. Cocaine abusers typically have overexpression of alpha-synuclein in dopamine neurons \cite{mash2003cocaine}. TH is the enzyme responsible for catalyzing the conversion of the amino acid L-tyrosine to L-3,4-dihydroxyphenylalanine (L-DOPA), which is a precursor for dopamine. Studies also suggested cocaine administration could increases TH enzyme activity \cite{vrana1993chronic}. STX1A is a nervous system-specific protein implicated in the docking of synaptic vesicles with the presynaptic plasma membrane. E-Cadherin is a type of cell adhesion molecule that is important in the formation of adherens junctions to bind cells with each other. Flotillin-1 and -2 are ubiquitously expressed, evolutionary conserved peripherally membrane-associated proteins. Flotillins are found to regulate membrane mobility of DAT.

The aforementioned  12 proteins  constitute the core network of cocaine addiction. Their mutual interactions in the network also indicate DAT is the most important node, which connects the upper and lower parts of the network. Network analysis shows that DAT has the highest degree (11) among all the nodes and is a hub of the core network. Additionally, the closeness centrality of DAT is as high as 1.000, which also suggests its full connection to all the other proteins in the core network. More importantly,  the betweenness centrality of DAT (0.470) is higher than any other nodes, suggesting other than the hub, DAT is also a critical bottleneck. In other words, DAT is a bridge of the network, and almost half of the interactions must be via DAT. If DAT is removed, the communication between the upper and lower parts of the core network will be essentially cut off.

The connections also reveal the importance of VMAT2 and alpha synuclein. Their degrees are both 8, and the closeness centrality values are both 0.786. They have connections to all nodes in the upper part of the network, and both have betweenness centrality values of 0.033,  forming shortcuts between other proteins. Another three proteins with positive betweenness centrality are D$_3$R, D$_2$R, and TH. Their betweenness centrality values are all 0.003, suggesting they play some roles as bottlenecks. For example, the shortest pathway between D$_1$R and D$_4$R is through D$_3$R or D$_2$R. Their degree and closeness centrality are 7 and all 0.733, respectively.

\subsection{Cross-target BA predictions}

As mentioned in Section \ref{section:ppi-network}, the global PPI network of cocaine addiction involves as many as 141 proteins, which not only play roles in cocaine addiction but also participate in other biological activities. A drug must be specific to its own target and not affect other protein functions to avoid  side effects. In this section, through ML/DL models, we systematically predict inhibitor BAs to analyze side effects and repurposing potential.

We collect inhibitor data from the ChEMBL database \cite{gaulton2012chembl}. We build ML models for 32 proteins that have sufficient inhibitor data to build such models. These models are used for drug repurposing and side effect studies. 

\begin{figure}[ht!]
	\centering
	\includegraphics[width=0.8\textwidth]{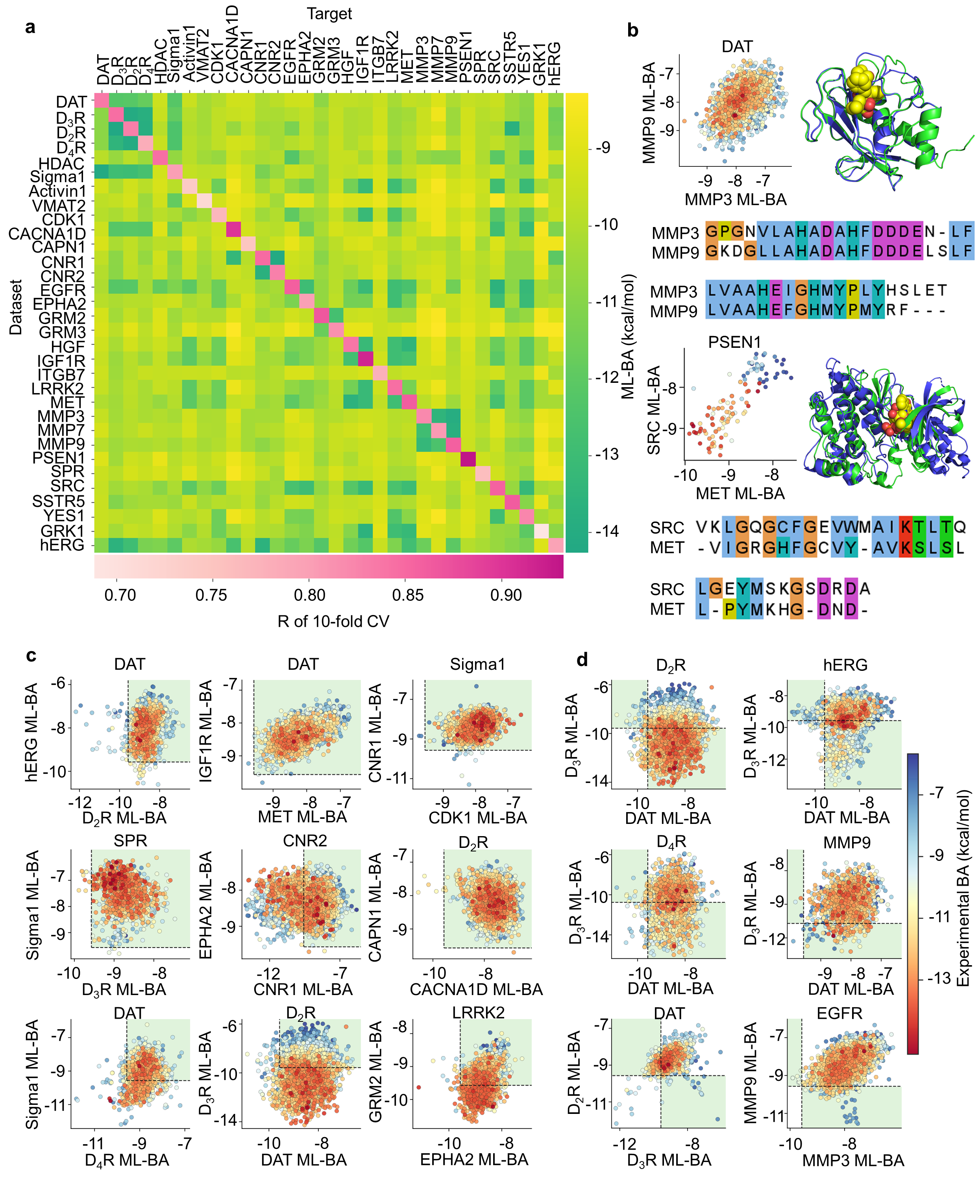}
	\caption{Cross-target BA prediction systematically suggesting side effects and repurposing potential. \textbf{a} The heatmap of cross-target BA prediction indicating the inhibitor specificity on each dataset. In each row, the diagonal element shows the Pearson correlation coefficients of 10-fold cross-validation (R of 10-fold CV) on the machine-learning predicted BAs (ML-BAs) of each dataset. Other elements represent the highest ML-BAs among the inhibitors in each dataset to other targets. \textbf{b} Two examples of positive cross-target BA correlation revealing binding-site similarity of MMP3 with MMP9 and MET with SPC. In each example, the ML-BA correlation plot of one dataset to the target pair, the 3D alignment of the two target proteins, and the 2D sequence alignment of their binding sites were given. The PDB IDs of the protein structures are 1B3D, 1GKC, 1R0P, and 1KSW for MMP3, MMP9, MET, and SPC, respectively. In each ML-BA correlation plot, the title is the name of the dataset. The colors of points represent the experimental BA to the designated target. The $x-$ and $y-$ axes indicate the ML-BA to two other proteins. \textbf{c} Nine typical examples of cross-target BA predictions of potential side effects. The first, second, and third rows exhibit the examples with substantial side effects of potent inhibitors to zero, one, and two targets, respectively. The green frames outline the optimal ranges without side effects to both the two targets ($x$ $>$ -9.57 kcal/mol and $y$ $>$ -9.57 kcal/mol). \textbf{d} Six typical examples of cross-target BA prediction suggesting repurposing potential. In these examples, some weak inhibitors to their designated targets are predicted to have high BAs (i.e., low BA values) to other proteins. The two green frames in each subplot outline the BA domains with repurposing potential, which represent compounds that have the potency to one target (BA values $<$ -9.57 kcal/mol) and do not show strong side effects to the other target (BA values $>$ -9.57 kcal/mol).}
	\label{fig:cross-prediction-heatmap}
\end{figure}

\subsubsection{Cross-target BA predictions}

Figure \ref{fig:cross-prediction-heatmap}a depicts the heatmap of cross-target BA predictions for 32 targets. 
Each diagonal element shows the Pearson correlation coefficient (R) of 10-fold cross-validation (CV) of the ML BA predictor (ML-BAs) for the corresponding protein dataset. Three out of 32 models have their R values higher than 0.90, showing excellent accuracy.  The R values of twenty one ML models are higher than 0.80. For example, the R value of the ML model for the DAT dataset of 2877 compounds is 0.84. Only one model's R value is less than 0.70 (i.e., R=0.69 for the GRK5 model). Therefore, these ML models are quite reliable.  

 In Figure \ref{fig:cross-prediction-heatmap}a, elements right to the diagonal in each row are the maximum BA values of the dataset of diagonal element predicted by the corresponding models. For example, element (1,2) is the maximum BA value of the DAT dataset (2877 compounds) predicted by the D$_3$R ML model. 
Elements below the diagonal in each column are the maximum BA values of other datasets predicted by the diagonal model. For example, element (2,1)  the maximum BA value of the D$_3$R dataset (4685 compounds) predicted by the DAT ML model.  

For a given drug candidate, its predicted high cross-target ML-BAs might suggest strong side effects. Among 992 cross-target screenings in Figure \ref{fig:cross-prediction-heatmap}a, there are 330 potential side effects judged by a threshold BA of -9.57 kcal/mol ($\rm K_i$=0.1 $\rm \mu$M), suggesting high side effects. 
Some side effects were due to highly similar targets, such as receptors D$_3$R, D$_2$R, and D$_4$R, the cannabinoid receptors  CNR1 and CNR2,  glutamate metabotropic receptors GRM2 and GRM3, as well as matrix metallopeptidases MMP3, MMP7, and MMP9. Their high sequence and structure similarities contribute their mutual side effects. However, mutual side effects are also found between seemingly unrelated proteins, such as DAT and Sigma1, etc.  

\subsubsection{Cross-target BA correlations revealing binding-site similarities} \label{section:similar_targets}

We next investigate the positive correlations between predicted ML-BAs. 
Figure \ref{fig:cross-prediction-heatmap}b exhibits two examples of cross-target BA correlations. The first example depicts compounds in the DAT dataset binding to targets MMP3 and MMP9, which play an important role in cocaine relapse \cite{smith2014synaptic}. The correlation plot reveals an R value of 0.48 between their predicted BAs.  The second and third plots are the 3D alignment of the proteins and 2D sequence alignment of their binding sites, respectively, which suggest these  proteins, and especially their binding sites, are highly similar, with a binding-site sequence identity as high as 64.9\%.
 
In addition to the targets from the same protein family leading to correlated BAs, we also found some seemingly unrelated proteins with correlated BAs, indicating their binding sites are similar as shown in Figure \ref{fig:cross-prediction-heatmap}b. Although tyrosine-protein kinase met (MET) and proto-oncogene tyrosine-protein kinase src (SRC) are not of the same family, they are both kinases. MET is a tyrosine kinase that transduces signals from the extracellular matrix into the cytoplasm by binding to hepatocyte growth factor ligand. SRC is a tyrosine kinase that is activated following the engagement of many different classes of cellular receptors. The alignment plots in Figure \ref{fig:cross-prediction-heatmap}b reveals the binding domains of MET and SRC having conserved 3D conformations with the 2D sequence identity 50.1\%. Thus, the R values between the BAs of MET and SRC datasets are as high as 0.82. More examples can be found in Figure S1 of the Supporting information.

\subsubsection{Systematic predictions of side effects and repurposing potentials}

One significant application of cross-target BA predictions is to evaluate side effects and repurposing potentials. Our basic idea is to systematically predict the BAs of the inhibitors of one target by using the ML models of datasets for other proteins. It is desirable for a drug candidate to be highly specific, i.e., having a high BA to its target, and low side effects, i.e., having very low BAs to all other human proteins. Moreover, if a drug candidate interacts weakly to its designated target but is predicted to be potent at another unintended protein, then it has  repurposing potential. Here, we carefully studied the 330 dataset-target pairs with potential mutual side effects in Figure \ref{fig:cross-prediction-heatmap}a. Figure \ref{fig:cross-prediction-heatmap}c and d depict some typical examples of our side-effect and repurposing detection through cross-target BA predictions. In each chart, three targets are involved: its designated target and two other potential side-effect targets.

Figure \ref{fig:cross-prediction-heatmap}c exemplifies side-effect predictions. The first row illustrates active  inhibitors having no serious side effects on either of two other targets. In the three plots, all active compounds, which are represented by red or even deep red points, are predicted to have low BAs (i.e., BA values $>$ -9.57 kcal/mol) for two other proteins. Therefore, we anticipated that these active inhibitors would not have strong side effects on  two other targets studied. The second row contains examples with predicted side effects on one of two targets. For instance, the second plot in this row shows that potent inhibitors of protein CNR2 are unlikely to bind to EPHA2. However, some of these potent inhibitors have strong predicted BAs ($\approx$ -12 kcal/mol) for CNR1. This potential side effect is expected, as CNR1 and CNR2 are similar cannabinoid receptors. The third row is the worst case in which side effects are predicted for both of the two other proteins. The most obvious cases are due to the kinship of the involved proteins, which are included in Figure S2 of the Supporting information. However, we also noticed that some inhibitors can still cause simultaneous side effects on unrelated targets, such as Sigma1 and D$_4$R in the first chart on this row.

Besides side-effect evaluation, our cross-target BA predictions could also suggest repurposing potential as shown in  Figure \ref{fig:cross-prediction-heatmap}d. In each subplot of Figure \ref{fig:cross-prediction-heatmap}d, some inactive inhibitors  are predicted to be potent  inhibitors of  other proteins. For instance, in the first chart, some inactive  D$_2$R inhibitors have high ML-BAs (BA values $<$ -9.57 kcal/mol) on DAT, suggesting these D$_2$R inhibitors are potential DAT inhibitors for further studies. In the second chart, some compounds inactive to hERG are predicted to be very potent to DAT, while some other inactive compounds are predicted to strongly block D$_3$R. In the fifth chart, some inactive DAT inhibitors are  predicted to be potent inhibitors of  D$_3$R or D$_2$R. More side-effect and repurposing examples are given in Figure S2 of the Supporting information.

\subsubsection{BA predictions targeting DAT and hERG}

Since DAT is the main target of cocaine and hERG is a critical side-effect target especially for neurotransmitter transport inhibitors and GPCR ligands, in this section, we focused on predicting the BAs of inhibitors from all the other 30 datasets  on DAT and hERG, which could evaluate their hERG side effects and repurposing potential against DAT.

Avoiding hERG side effects is a priority for all  drugs. As shown in Figure S3, most of the inhibitors in many datasets have predicted hERG BA values higher than -8.18 kcal/mol ($\rm K_i$=1 $\rm \mu$M), which suggests no serious hERG side effect. Especially for datasets GRM3, LRRK2, and SPR, almost all the compounds in these datasets  weakly bind to hERG with the BA values higher than -8.18 kcal/mol. However, a large number 
molecules in datasets D$_2$R, D$_3$R, and D$_4$R were predicted to have serious hERG side effects.

In searching the repurposing potential for DAT compounds, inactive inhibitors in datasets D$_2$R, D$_3$R, D$_4$R, EGFR, MET, and Sigma1 with experimental BA values $>$ -9.57 kcal/mol are predicted to be potent at DAT with their BA values lower than -9.57 kcal/mol for DAT. Some of these compounds possess a low potential for hERG side effects as shown in Figure S3. Of particular note is the Sigma1 dataset containing 44 inactive inhibitors but predicted to strongly inhibit DAT with ML-BA values lower than -9.57 kcal/mol. Additionally, 11 of the 44 compounds were predicted to have weak hERG side effects with BA values higher than -8.18 kcal/mol. Therefore, according to our predictions, these 11 compounds could potentially be repurposed for inhibiting DAT without strong hERG side effects. However, to be qualified as nearly optimal lead compounds, further  screenings for other side effects, ADMET properties, and synthesizability are indispensable. 

\subsection{Druggable property screening}

\begin{figure}[ht!]
	\centering
	\includegraphics[width=1.0\textwidth]{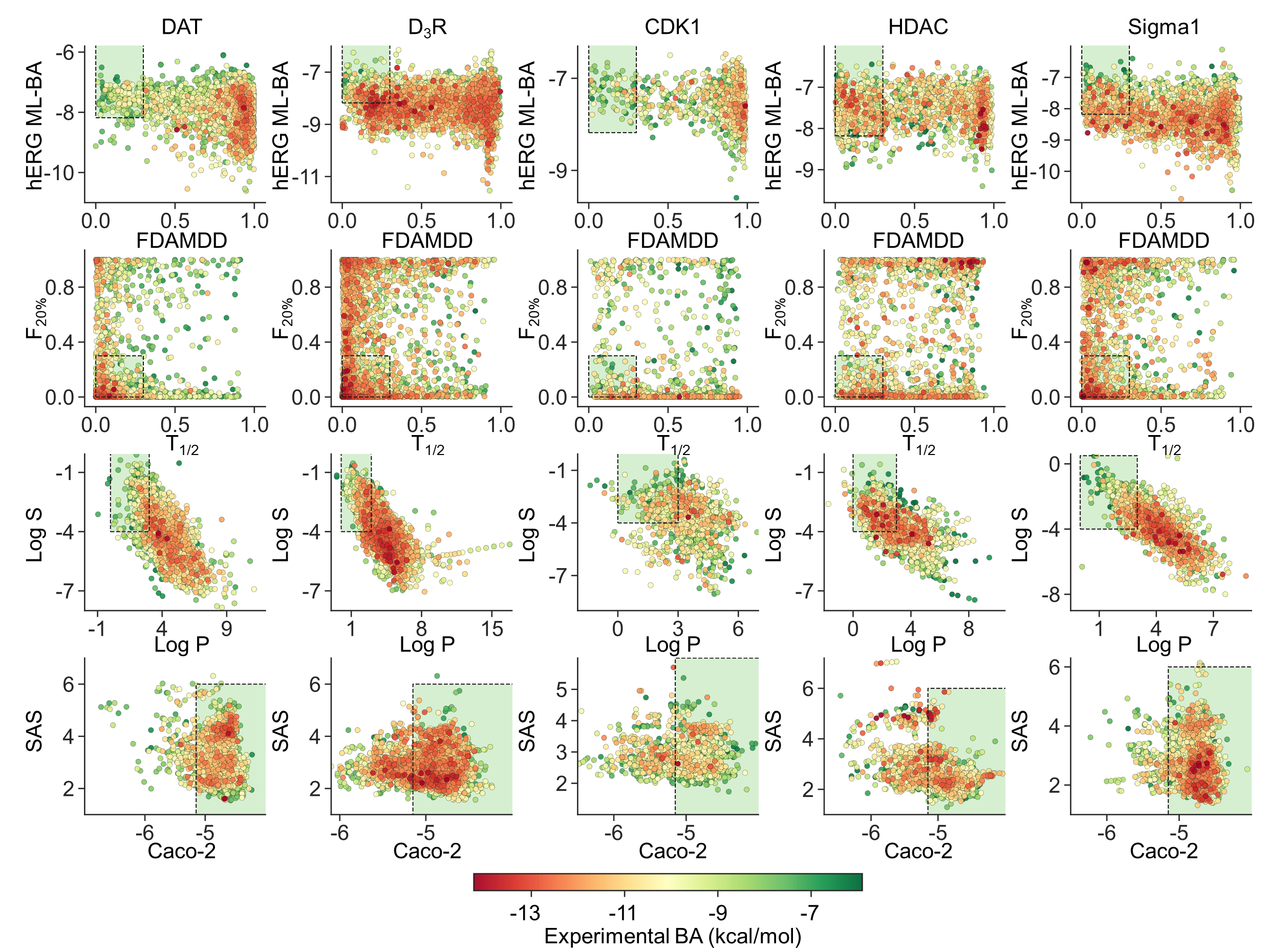}
	\caption{Druggable property screening based on   ADMET properties, synthesizability, and hERG side-effects to compounds from five critical protein datasets: DAT, D$_3$R, CDK1, HDAC, and Sigma1. The colors of points represent the experimental BAs to these targets. The $x-$ and $y-$axes show   predicted ADMET properties, synthesizability, or hERG side-effects. Green frames outline the optimal ranges of these properties and side-effects.}
	\label{fig:admet-dist}
\end{figure}

In this section, we performed systematic screenings on ADMET properties, synthesizability, and hERG side effects. Figure \ref{fig:admet-dist} illustrates the example screening implemented on the datasets of five proteins: DAT, D$_3$R, CDK1, HDAC, and Sigma1, which play essential roles in cocaine addiction. The optimal ranges of ADMET properties and synthesizability are provided in Table \ref{tab:property-optimal}, while the BA value $>$ -8.18 kcal/mol is applied as the required range for exempting hERG side effects.

Two critical properties for potential drug candidates are FDA maximum recommended daily dose (FDAMDDs) and the BA for hERG (hERG\_BA), representing the potential for toxicity and hERG side effects, respectively. The first row of Figure \ref{fig:admet-dist} depicts the distributions of these two properties of inhibitors from the five critical datasets. The green frames are the optimal domains of the above two properties. The colors of points represent  experimental BA values for targets. According to this screening, all the five datasets contain sufficient compounds with optimal toxicity and hERG side effects. However, for the CDK1 and DAT datasets, the optimal domains of toxicity and hERG side effects only contain very few potent inhibitors. This suggests ADMET properties and side effects must be  considered before a new compound is synthesized. 

The second row of Figure \ref{fig:admet-dist} illustrates the screening based on important absorption properties $\rm T_{1/2}$ (half-life) and $\rm F_{20\%}$ (bioavailability 20\%). All the five plots in the second row reveal that the optimal domain of the $\rm T_{1/2}$ and $\rm F_{20\%}$ is only a small fraction of chemical space. However,  for all these five datasets, the small optimal domain does indeed contain some potent inhibitors.

The third row of Figure \ref{fig:admet-dist} displays the log P and log S screening. Log P and log S relate to the distribution of chemicals in human bodies. For all  five targets, only a small portion of potent inhibitors can be found in the optimal domain, suggesting a huge waste of resources in early studies. Notably, there is an obvious line in the second subplot of this row, which is very unusual under natural conditions. This obvious line is probably caused by intended optimization to improve log P.

The last row of Figure \ref{fig:admet-dist} depicts the caco-2 and SAS screening. Caco-2 represents the cell permeability of compounds, while SAS describes how hard a compound is to synthesize. These plots indicate almost all the compounds from the five datasets are not hard to synthesize, and about half of the compounds have good cell permeability. More importantly, many potent inhibitors are also in the optimal domain.

\section{Discussion}

\subsection{Side-effect predictions of existing experimental medications}
According to literature  \cite{newman2021new}, some experimental medications are being investigated to treat cocaine addiction. Here we used our proteome-informed ML models to predict the side effects of these experimental medications. Ibogaine is a naturally occurring psychoactive substance that may have anti-addiction properties, and its docking structure and some experimental/predicted BAs  are  shown in Figure \ref{fig:comps-struct}c. All the 2D structures of existing experimental medications discussed in this work and their experimental or predicted BA values are available in Figures S4 and S5 of the Supporting information.


\subsubsection{Atypical DAT inhibitors}

A typical DAT inhibitor such as cocaine binds and stabilizes an outward-facing conformation of DAT. However, an atypical DAT inhibitor stabilizes an inward-facing conformation of DAT upon its binding, which makes the binding of  cocaine difficult \cite{reith2015behavioral}. Therefore, even with a weaker BA on DAT than cocaine, the pretreatment by an atypical inhibitor could still prevent DAT from being blocked by cocaine. More importantly, compared to typical DAT inhibitors such as cocaine, atypical DAT inhibitors increase the accessibility of residues in the cytoplasmic substrate permeation pathway and do not induce a cocaine-like behavior or addiction \cite{reith2015behavioral}.

\paragraph{Ibogaine}

\begin{figure}[ht!]
	\centering
	\includegraphics[width=0.8\textwidth]{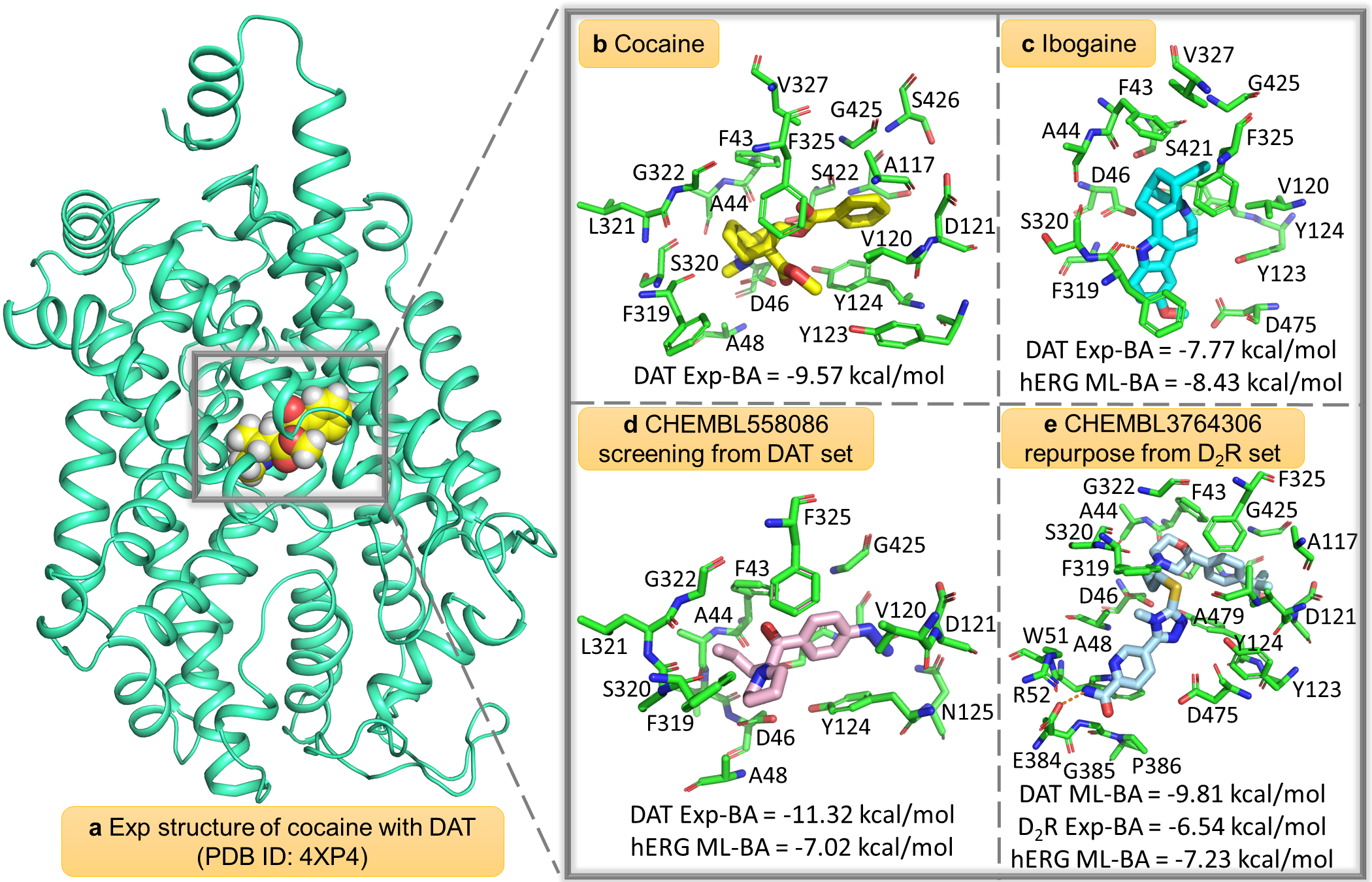}
	\caption{Cocaine as well as some examples of existing experimental medications and nearly optimal lead compounds from our ML-based screening and repurposing for treating cocaine addiction. The cocaine binding conformation is from the crystal structure  (PDB ID 4XP4) \cite{wang2015neurotransmitter}, while others are obtained from molecular docking.}
	\label{fig:comps-struct}
\end{figure}

Ibogaine is a natural psychoactive substance extracted from the plants in the  Apocynaceae family such as Tabernanthe iboga, Voacanga africana, and Tabernaemontana undulata (see Figure \ref{fig:comps-struct}b). Ibogaine was originally used in African spiritual ceremonies. However, its anti-addictive properties were accidentally discovered in 1962. Since then, ibogaine has been tested to treat SUD, especially for cocaine addiction. Now it is already approved for clinical use in the Netherlands, Canada, and Mexico.

Ibogaine inhibits both DAT and SERT with $\rm IC_{50}$ values of 4.0 $\rm \mu M$ and 0.59 $\rm \mu M$, respectively \cite{efange1998modified}. More importantly, Ibogaine is an atypical inhibitor of DAT and SERT, and has a potential for treating cocaine addiction. However, its severe side effects and related deaths are of serious concern. Between 1990 and 2008, a total of 19 fatalities associated with the ingestion of Ibogaine were reported, and 6 of these fatalities were caused by acute heart failure or cardiopulmonary arrest \cite{koenig2015anti}.  
Our model predicted a moderate BA of Ibogaine to hERG with a BA value -8.43 kcal/mol, which suggests a risk for Ibogaine  to incur heart issues.  Additionally, with predicted BA values of -9.71 kcal/mol, -9.54 kcal/mol, -9.46 kcal/mol, and -9.43 kcal/mol to proteins YES1, LRRK2, GRM2, and FER, respectively, our models anticipated high risks of side effects on these proteins, especially for YES1 and FER associated with severe diseases sarcoma and acute myeloid leukemia, respectively.

The docking structure of Ibogaine to DAT in Figure \ref{fig:comps-struct}c reveals that, just like the crystal structure of cocaine with DAT in Figure \ref{fig:comps-struct}b, Ibogaine binds to DAT mainly through hydrophobic interactions. Strong hydrogen interactions are absent. 

\paragraph{Modafinil and its analogs}

Modafinil is a functional stimulant targeting DAT and NET \cite{ballon2006systematic}, which has been approved to treat excessive sleepiness, such as narcolepsy
or idiopathic hypersomnia. It was also reported that modafinil exerts some effects on cocaine reward and reinforcement \cite{ballon2006systematic}. As a result, its potential utility to establish abstinence from cocaine addiction was tested in a phase I clinical trial, which reported a cocaine-blunting
effect \cite{dackis2003modafinil}. As an atypical DAT inhibitor,    even with a lower DAT affinity  ($K_i$ = 8.16 $\rm \mu M$ \cite{newman2021new}) than cocaine, Modafinil could still prevent DAT from be blocked by cocaine.

Using the proteome-informed ML models, we also predicted the side effects of Modafinil on other proteins in our cocaine-addiction PPI network. Consistent with the fact that Modafinil is an approved drug, its predicted BA values to these side-effect targets are all higher than -9 kcal/mol, and among them, 65.7\% are higher than -8 kcal/mol. Especially important is that  Modafinil only has a very weak binding to hERG (-6.96 kcal/mol). 

Since Modafinil is a potential treatment for cocaine addiction, many of its analogs have been developed and tested. The first modafinil analog is JJC8-016 \cite{zhang2017novel}, since it is a potent atypical DAT inhibitor ($K_i$ = 0.116 $\rm \mu M$), and its pretreatment inhibits cocaine enhanced locomotion, cocaine self-administration, and cocaine-induced reinstatement of drug-seeking behavior. However, an experimental study found that it may block the hERG channel \cite{tunstall2018atypical}. Our prediction to JJC8-016 also showed a BA value of -10.34 kcal/mol for hERG, which suggests serious potential cardiotoxicity.

Another series of modafinil analogs are JJC8-088, JJC8-089, JJC8-091 \cite{newman2019translating}, and RDS3-094 \cite{slack2019structure}. Their $K_i$s to DAT are 0.0026 $\rm \mu M$, 0.0378 $\rm \mu M$, 0.23 $\rm \mu M$, and 0.0231 $\rm \mu M$, respectively. Although  JJC8-091 is not as potent as  JJC8-088 and JJC8-089, its efficiency in blocking cocaine effects is the best among these three. The possible reason is that JJC8-091 prefers an inward-facing conformation of DAT, and thus exhibits a stronger atypical inhibition \cite{newman2019translating}. As a result, JJC8-091 is under further investigation. RDS3-094 is a newly developed modafinil analog in 2020. With similar chemical structures, JJC8-088, JJC8-089, JJC8-091, and RDS3-094 were predicted to have similar side-effects. In our prediction, they all strongly bind to targets Sigma1 (BA values: -10.06 kcal/mol, -10.15 kcal/mol, -10.16 kcal/mol, and -9.99 kcal/mol) and YES1 (BA values: -9.15 kcal/mol, -9.31 kcal/mol, -9.25 kcal/mol, and -9.14 kcal/mol), and YES1 is a risky target associated with sarcoma. Additionally, JJC8-088 strongly interacts with SSTR5, which relates to pituitary adenoma. In contrast, their hERG side effects are not very obvious with BA values of -8.59 kcal/mol, -8.48 kcal/mol, -8.03 kcal/mol, and -8.08 kcal/mol. Interestingly, these four modafinil analogs were also predicted to inhibit $D_3$R (-9.51 kcal/mol, -9.07 kcal/mol, -9.10 kcal/mol, and -9.04 kcal/mol), which suggests they may assuage cocaine addiction via multiple mechanisms.

\paragraph{Benztropine and its analogs}

Benzatropine is a medication for movement disorders including dystonia and parkinsonism. Benzatropine and its analogs are being studied to repurpose against cocaine addiction. Although benztropine pretreatment failed to significantly affect responses to acute cocaine administration \cite{penetar2006benztropine}, its analogs are still worth investigating. For instance, JHW007, with a $K_i$ of 0.0253 $\rm \mu M$, shows antagonism of behaviors produced by cocaine or methamphetamine across numerous animal models \cite{velazquez2013atypical}. However, serious side effects of JHW007 were predicted on targets Sigma1, GRM2, and YES1 with affinities of -9.46 kcal/mol, -9.30 kcal/mol, and -9.28 kcal/mol, respectively. The abnormality of YES1 could lead to sarcoma.

\paragraph{Rimcazole and its analogs}

Rimcazole was originally designed as a potential antipsychotic. However, trials indicate rimcazole is not effective in this application. Instead, Rimcazole and its analogs can reduce the effects of cocaine \cite{katz2003behavioral},
specifically through binding to DAT. Rimcazole was reported to have a $\rm K_i$ of 0.0977 $\rm \mu M$ to DAT \cite{newman2021new}. Since it is still a drug candidate under investigation, the side effects remain a major concern. Our results indicate potentially serious side effects on targets $\rm D_4R$ and YES1 with BA values of -9.60 kcal/mol and -9.24 kcal/mol, respectively. 

Rimcazole analog GBR12909 (vanoxerine) is a potent atypical DAT inhibitor with $K_i$ = 0.00177 $\rm \mu M$. Therefore, it was advanced to phase I clinical trials. However, a failure was reported due to rate-dependent corrected QT (QTc) elongation in healthy subjects \cite{newman2021new}. This heart-related side effect was supported by the BA value for hERG, which was predicted to be as low as -9.47 kcal/mol. Another potential side effect is from Sigma1, with a predicted BA value of -9.43 kcal/mol.

\paragraph{Dextroamphetamine}

Dextroamphetamine is an approved drug prescribed for treating attention deficit hyperactivity disorder (ADHD) and narcolepsy. Because it has potent atypical antagonism of DAT with a $K_i$ of 0.109 $\rm \mu M$ \cite{arunotayanun2013analysis}, it is being reexamined to treat cocaine addiction. In clinical trials, Dextroamphetamine has demonstrated significant promise to reduce cocaine effects \cite{grabowski2001dextroamphetamine}. Consistent with the fact Dextroamphetamine is already approved, the predicted side effects to other targets are negligible in our models. For example, the predicted BA value to hERG is -7.09 kcal/mol.

\subsubsection{D$_3$R antagonists/partial agonists}

\paragraph{D$_3$R antagonists}
The D$_3$R antagonist SB277011A  is studied for cocaine addiction. It can reduce cocaine-enhanced brain-stimulation reward, suppress cocaine-conditioned place preference, and attenuate cocaine-primed reinstatement behaviors in rats \cite{guerrero2019modulation}. However, SB277011A is ineffective when cocaine is available for self-administration  \cite{di2003attenuation}. Considering side effects, SB277011A has approximately 100-fold selectivity for D$_3$R ($\rm K_i$ = 0.0112 $\rm \mu$M) over D$_2$R ($\rm K_i$ = 555 $\rm \mu$M). We also predicted its other side effects to more dopamine-related targets, and  high predicted affinities occur on targets YES1, SSTR5, HGF, and LRRK2 (BA values -9.71 kcal/mol, -9.39 kcal/mol, -9.23 kcal/mol, and -9.17 kcal/mol, respectively). Among them, the abnormality of YES1 and SStR5 could trigger serious consequences since YES1 and SSTR5 are involved in sarcoma and prolactin-secreting pituitary adenoma.

NGB2904 is another  widely investigated D$_3$R antagonist. In rats, NGB2904 reduces progressive ratio breakpoints for cocaine, inhibits both cocaine- and cue-primed reinstatement of cocaine-seeking, and prevents cocaine-enhanced brain-stimulation reward. However, its reward-attenuating effects can be overcome by higher doses of cocaine \cite{xi2007pharmacological}. In our side-effect prediction, a potential side effect on hERG was detected with the predicted BA value of -9.20 kcal/mol, suggesting the risk of causing heart issues. Serious side effects were also predicted on multiple targets, including YES1, activin1, SSTR5, CNR1, sigma1 with BA values s of -9.64 kcal/mol, -9.60 kcal/mol, -9.57 kcal/mol, -9.53 kcal/mol, and -9.47 kcal/mol, respectively, again raising serious concerns over sarcoma and prolactin-secreting pituitary adenoma. However, interestingly, NGB2904 was also predicted to potently inhibit DAT, suggesting NGB2904 may interact with multiple targets ($D_3R$ and DAT) to affect cocaine addiction behaviors.


PG01037 is derived from NGB2904. Similar to NGB2904, PG01037 reduces progressive ratio breakpoints for cocaine, inhibits both cocaine- and cue-primed reinstatement of cocaine-seeking, and prevents cocaine-enhanced brain-stimulation reward, but fails to alter cocaine self-administration \cite{higley2011pg01037}. PG01037 was predicted to have weaker side effects than NGB2904, especially as predicted BA values to hERG and YES1 are increased to -8.25 kcal/mol and -9.11 kcal/mol. Moreover, similar to NGB2904, PG01037 has a low predicted BA value to DAT (-9.56 kcal/mol).

\paragraph{D$_3$R partial agonists}

Many D$_3$R partial agonists are also under investigation now. VK4-40 and VK4-116 are D3R partial agonists designed within the last five years. They were developed to improve blood–brain barrier (BBB) penetration of previous D$_3$R partial agonists. Compared with VK4-40, VK4-116 can reduce cocaine-enhanced heart rate and blood pressure as well \cite{shaik2019investigation}. As for side effects, VK4-40 and VK4-116 exhibit high selectivity for D$_3$R (305-fold and 1735-fold more selective for D$_3$R over D$_2$R, respectively). However, our predictions show that they have potentially strong side effects on targets YES1 (BA values -10.07 kcal/mol and -10.27 kcal/mol), SSTR5 (BA values -9.44 kcal/mol and -9.66 kcal/mol), and LRRK2 (BA values -9.28 kcal/mol and -9.30 kcal/mol). 

The partial D$_3$R agonist CJB090 more effectively attenuated psychostimulant reward than PG01037 in rats  \cite{orio2010dopamine}. We predicted its side effects on other targets, revealing the BA to YES1 (-9.27 kcal/mol) is weaker than that of VK4-40 and VK4-116. However, the potential side effect on SSTR5 is still somewhat high (BA value -9.50 kcal/mol).

Cariprazine is an atypical antipsychotic already on the market. It acts primarily as a D$_3$R and D$_2$R partial agonist, with high selectivity for D$_3$R, and it is also being studied for treating cocaine addiction. Cariprazine exhibits a reduction of cocaine intake under an FR1 schedule and reinstatement of cocaine-seeking in rats \cite{roman2013cariprazine}. It is only predicted to have a high side effect (-9.82 kcal/mol) on  Sigma1, and  Sigma1 is not currently known to be related to serious diseases.

Other experimental D$_3$R partial agonists include BP-897, RGH-237, and GSK598809. BP-897 reduces conditioned locomotor activity to cocaine, but it fails when cocaine is self-administrated \cite{cervo2003cocaine}. As for side effects, the $\rm K_i$ assays indicate BP-897 is about 70-fold more selective for D$_3$R over D$_2$R. Other side effects were predicted via our proteome-informed models. Potential serious side effects were found for targets SSTR5 and YES1 with predicted BA values of -9.73 kcal/mol and -9.38 kcal/mol. RGH-237 was also predicted to have some high affinities to SSTR5 and YES1 (BA values -9.79 kcal/mol and -9.65 kcal/mol). Notably, our predictions suggest GSK598809 strongly binds to targets YES1 and CNR1 with BA values  of -10.15 kcal/mol and -9.94 kcal/mol, especially a -10.15 kcal/mol affinity to YES1 that may  represent a danger for serious side effects, since YES1 relates to sarcoma.


\subsection{Nearly optimal leads from our systematic screening and repurposing}

In this section, using our proteome-informed ML models, we mimicked the processes of screening or repurposing lead compounds. We applied as many as 38 criteria in our systematic screening and repurposing. These   criteria include BA to the designated target, six ADMET properties in Table \ref{tab:property-optimal}, synthesizability, as well as the side effects to hERG and   other 30 proteins relating to cocaine addiction in Figure \ref{fig:cross-prediction-heatmap}a. 

Figure \ref{fig:comps-struct}d exemplifies our systematic screening. In this example, we screened nearly optimal lead compounds from known DAT inhibitors in the ChEMBL database. We only accepted the ones with experimental DAT BA values lower than -9.54 kcal/mol (K$\rm _i$ $<$ 0.1 $\rm \mu M$),  predicted hERG BA values higher  than -8.18 kcal/mol (K$\rm _i$ $>$ 1 $\rm \mu M$), predicted BA values on other proteins higher than -9.54 kcal/mol, as well as excellent predicted ADMET properties and synthesizability. As a result, screened compounds have high potency to DAT, low side effects on hERG and other targets, and satisfy standards for druggable properties. Also, they are easy to synthesize. Compound CHEMBL558086 in Figure \ref{fig:comps-struct}d was the only optimal lead compound left from our ML-base screening. It has a stronger binding to DAT than to cocaine (Experimental BA value = -11.32 kcal/mol) and weak side effects on all the other proteins, such as a predicted hERG BA value of -7.02 kcal/mol. The predicted ADMET properties and synthesizability are also in the excellent ranges of Table \ref{tab:property-optimal}. 

Figure \ref{fig:comps-struct}(e) shows an example compound from our systematic ML-based repurposing. In this example, we hope to repurpose inhibitors of PPI-informed proteins from ChEMBL to target DAT. For this purposing, we searched the compounds with weak affinity to their designated targets but potent binding to DAT by our ML predictions. At the same time, they must have low side effects on other proteins such as hERG and good druggable properties. Our criteria are the same as those for the screening described above, but input compounds are from other datasets. As a result, the most potent DAT inhibitor from our systematic repurposing is ChEMBL3764304 from the D$_2$R dataset. It is very weak to D$_2$R, but it is predicted to be effective to DAT with a BA value of -9.81 kcal/mol.
Moreover, its side effects are low. For example, its predicted hERG BA value is only -7.23 kcal/mol.


\section{Methods}
 
\subsection{The PPI network and analysis} 

In this work, our PPI networks related to cocaine addiction were obtained from the String website (https://string-db.org/). The network analysis and visualization were implemented via Cytoscape 3.8.2 \cite{shannon2003cytoscape}. In the network analysis, we considered three indexes: degree, between centrality, and closeness centrality. The degree of a node is the number of edges. A node with a high degree represents a hub node having many neighbors. Between centrality of a node is defined as the proportion of the number of the shortest paths via it to the number of all the shortest paths in the network, which quantifies the frequency of a node forms the shortest paths between two other nodes. A node with a high between centrality is always a bottleneck of the network and dramatically influences the pathways among other nodes. Additionally, the closeness centrality of a node, which is defined as the average length of the shortest paths between the node and all other nodes, measures its centrality in the network. A node with higher closeness centrality is closer to the center of the network.

\subsection{Molecular fingerprints} 

Molecular fingerprints are property profiles of a molecule, usually in the form of vectors with each vector element indicating the existence, the degree, or
the frequency of one particular structure characteristic. They can be used as features for ML/DL models. In this work, the latent-vector fingerprint (LV-FP) and traditional 2D fingerprints (2D-FPs) were applied.

\subsubsection{The seq2seq LV-FP}
The seq2seq model is a DL autoencoder architecture originated from natural language processing. It has already demonstrated as a breakthrough success in English-French translation and conversational modeling. The basic scheme of the seq2seq model is to map an input sequence to a fixed-sized latent vector in the latent space using a gated recurrent unit (GRU) \cite{cho2014learning} or a long short-term memory (LSTM) network \cite{hochreiter1997long}, and then map the vector to a target sequence with another GRU or LSTM network. 

In our study, input and output sequences are both SMILES strings -- a one-dimensional ``language'' of chemical structures. Using near 104 million of molecules, our autoencoder model is trained to have a high reconstruction ratio between input and output smiles strings so that the latent vectors contain faithful information of the chemical structures. Thus, we adopted these latent vectors as LV-FP to represent compounds.

The seq2seq model and LV-FPs were realized by our in-house source code. We applied bidirectional LSTMs as the encoder. The generated LV-FP has a dimension of 512. The structure of our seq2seq model is illustrated in Figure S7.

\subsubsection{Traditional 2D-FPs}
Besides LV-FP, 2D-FP based predictors were also adopted in our work. The predictions from 2D-FPs were combined with those from LV-FP by consensus (averages of their prediction values) to further enhance predictive power. According to our previous tests \cite{gao20202d}, ECFP4 \cite{rogers2010extended}, Estate1\cite{hall1995electrotopological}, and Estate2 \cite{hall1995electrotopological} fingerprints perform best on BA prediction tasks. Thus these three 2D fingerprints were also considered in this work. We employed the RDKit software (version 2018.09.3) \cite{landrum2006rdkit} to generate 2D-FPs from SMILES strings.  

\subsection{BA predictions}

To achieve fast and robust BA predictions, we performed gradient boosting decision tree (GBDT) regressors and classifiers using GradientBoostingRegressor and GradientBoostingClassifier modules in scikit-learn (version 0.20.1) \cite{pedregosa2011scikit}. The hyperparameters were tuned according to 10-fold cross-validation tested on different datasets.

For regression tasks, the evaluation criteria are the square of Pearson correlation coefficient ($\rm R^2$), and root mean square error (RMSE). For classification tasks, the evaluation criteria are accuracy, F score, sensitivity, and specificity. 

\subsection{Predictions of ADMET properties and synthesizability}

ADMET properties and synthesizability are also critical factors for drug design and lead optimizations. In this work, we focused on seven different indexes of ADMET and synthesizability: FDA maximum recommended daily dose (FDAMDD), log P, log S, half-life ($\rm T_{1/2}$), caco-2 permeability (caco-2), human oral bioavailability 20\% ($\rm F_{20\%}$), Synthetic accessibility score (SAS). Among them, FDAMDD is the maximum daily dose recommended by the FDA, which estimates the toxic dose threshold of a compound in humans. Log P is the logarithm of the n-octanol/water distribution coefficient. Log S is the logarithm of the aqueous solubility value. Log P and log S relate to the distribution of chemicals in human bodies \cite{di2006biological}. The half-life time of a drug indicates the length of time that the drug effect could persist in an individual. The value of $\rm T_{1/2}$ here represents the probability of half-life time shorter than 3 hours. Caco-2 estimates in-vivo drug permeability. $\rm F_{20\%}$ measures the fraction of the initial dose of a drug that successfully reaches either the site of action or the bodily fluid domain from which the drug intended targets have unimpeded access. Last but not least, SAS represents synthesizability, namely the ease of synthesis of a drug-like molecule. We adopted ADMETlab 2.0 (https://admetmesh.scbdd.com/) \cite{xiong2021admetlab} to predict these seven properties. Its document also provides optimal ranges for these ADMET properties, as shown in Table \ref{tab:property-optimal}.

\begin{table}
	\centering
	\begin{tabular}{c|c}		
		\hline
		
		Property & Optimal range   \\ 	
		\hline
		
		FDAMDD & Excellent: 0-0.3; medium: 0.3-0.7; poor: 0.7-1.0  \\
		
		$\rm F_{20\%}$ & Excellent: 0-0.3; medium: 0.3-0.7; poor: 0.7-1.0  \\
		Log P & The proper range: 0-3 log mol/L \\
		Log S & The proper range: -4-0.5 log mol/L \\
		
		$\rm T_{1/2}$ & Excellent: 0-0.3; medium: 0.3-0.7; poor: 0.7-1.0  \\
		Caco-2 & The proper range: $>$-5.15 \\
		SAS & The proper range: $<$6 \\
		\hline

	\end{tabular}
	\caption{The optimal ranges of the ADMET properties and synthesizability considered in this work: FDAMDD, $\rm F_{20\%}$, log P, log S, $\rm T_{1/2}$, caco-2, and SAS.}
	\label{tab:property-optimal}
\end{table}

\subsection{Cocaine-addiction related datasets}

This work studied 36 datasets for 32 different protein targets. The data were collected from literature \cite{fant2019toward, warszycki2021pharm} and ChEMBL database \cite{gaulton2012chembl}. These datasets are summarized in Table S1.

For each dataset, we included all the compounds with $\rm K_i$ or $\rm IC_50$ values but removed redundant ones. Suggested by Kalliokoski et al. \cite{kalliokoski2013comparability}, the $\rm IC_{50}$ values were approximately converted to $\rm K_i$ by $\rm K_i$ = $\rm IC_{50}$/2. The label we used for training and testing is the binding affinity (1.3633 $\times \rm log_{10} K_i$).

In the benchmark classification tasks, to define a distinct boundary between active and inactive, we remove the compounds between active and inactive following the strategies from the literature: for the DAT and hERG datasets, the compounds with $\rm pK_i$s or $\rm pIC_{50}$s value between 5 and 6 were excluded; for the D$_2$R dataset, the compounds with $\rm pK_i$s or $\rm pIC_{50}$s value between 6 and 7 were removed. 

\subsection{Our prediction performances on benchmark datasets}

\begin{figure}[ht!]
	\centering
	\includegraphics[width=0.9\textwidth]{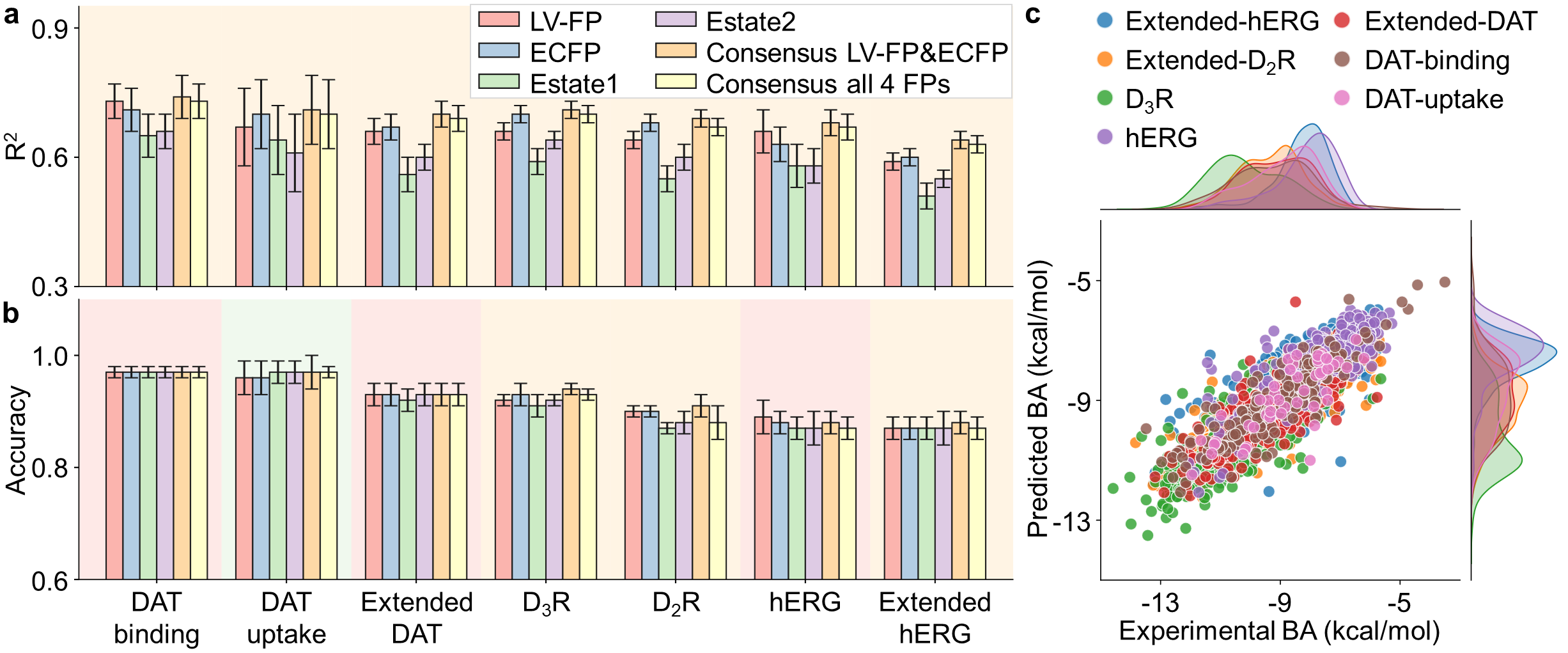}
	\caption{Performance of our predictive models on some benchmark datasets relating to cocaine addiction.}
	\label{fig:prediction-performance}
\end{figure}

Some benchmark datasets relating to cocaine addiction were already studied in terms of regression or classification tasks\cite{fant2019toward,warszycki2021pharm}. On these benchmark datasets, our predictors, especially consensus ones, exhibit high prediction power (see Figure \ref{fig:prediction-performance}). On regression tasks, we can achieve R values over 0.8 ($\rm R^2$ $>$ 0.64) on all the datasets except extended hERG. The R on the extended hERG dataset is close to 0.8 (R = 0.77, $\rm R^2$ = 0.59). These performances are better than that in the existing reports \cite{fant2019toward}. More detailed comparisons are shown in Tables S2-S13 in the Supporting information.

\section*{Key Points}
\begin{itemize}
	
\item This work addresses the urgent need for effective drugs to deal with cocaine addiction. There are no FDA-approved drugs for cocaine dependence at present.  

\item We propose a proteome-informed machine learning platform, which results in 141 drug targets for cocaine dependence. Using our autoencoder trained from over 104 million molecules, we built 32 machine-learning models for the targets with enough existing training data.

\item We perform cross-target analysis of over 60,000 drug candidates or experimental drugs to predict their side effects and repurposing potentials for treating cocaine addiction.

\item We further screen the absorption, distribution, metabolism, excretion, and toxicity (ADMET) properties of these candidates and experimental drugs.

\item Our platform reveals that essentially all existing drug candidates, including dozens of experimental drugs, fail to pass our cross-target and ADMET screenings, which explains why there are no FDA-approved anti-cocaine addiction drugs. Nonetheless, we have identified two nearly optimal leads for further optimization. Our work opens a novel, proteome-informed machine learning direction for drug discovery. 
	
\end{itemize}

\section*{Data  availability}

 The 36 cocaine-addiction related datasets studied in this work are available at: \\
https://weilab.math.msu.edu/DataLibrary/2D/.

\section*{Model  availability}
Our source code and trained autoencoder model for LV-FP generation can be found at\\ https://github.com/KfGao2021/antoencoder-v01.

\section*{Acknowledgment}
This work was supported in part by NIH grant  GM126189, NSF Grants DMS-2052983,  DMS-1761320, and IIS-1900473,  NASA 80NSSC21M0023,  MSU Foundation, Michigan Economic Development Corporation,  George Mason University award PD45722,  Bristol-Myers Squibb 65109, and Pfizer.

\section*{References}

\renewcommand\refname{}
%
\bibliographystyle{unsrt}
\bibliography{refs}

\begin{thebibliography}{10}

\bibitem{newman2021new}
Amy~Hauck Newman, Therese Ku, Chloe~J Jordan, Alessandro Bonifazi, and
  Zheng-Xiong Xi.
\newblock New drugs, old targets: Tweaking the dopamine system to treat
  psychostimulant use disorders.
\newblock {\em Annual Review of Pharmacology and Toxicology}, 61:609--628,
  2021.

\bibitem{cheng2015insights}
Mary~Hongying Cheng, Ethan Block, Feizhuo Hu, Murat~Can Cobanoglu, Alexander
  Sorkin, and Ivet Bahar.
\newblock Insights into the modulation of dopamine transporter function by
  amphetamine, orphenadrine, and cocaine binding.
\newblock {\em Frontiers in neurology}, 6:134, 2015.

\bibitem{matuskey2014dopamine}
David Matuskey, Jean-Dominique Gallezot, Brian Pittman, Wendol Williams, Jane
  Wanyiri, Edward Gaiser, Dianne~E Lee, Jonas Hannestad, Keunpoong Lim,
  Minq-Qiang Zheng, et~al.
\newblock Dopamine d3 receptor alterations in cocaine-dependent humans imaged
  with [11c](+) phno.
\newblock {\em Drug and alcohol dependence}, 139:100--105, 2014.

\bibitem{baik2013dopamine}
Ja-Hyun Baik.
\newblock Dopamine signaling in reward-related behaviors.
\newblock {\em Frontiers in neural circuits}, 7:152, 2013.

\bibitem{paul1992d1}
ML~Paul, AM~Graybiel, JC~David, and HA~Robertson.
\newblock D1-like and d2-like dopamine receptors synergistically activate
  rotation and c-fos expression in the dopamine-depleted striatum in a rat
  model of parkinson's disease.
\newblock {\em Journal of Neuroscience}, 12(10):3729--3742, 1992.

\bibitem{RAMANATHAN2018}
Sudarshini Ramanathan and Sarosh~R. Irani.
\newblock Autoantibody-mediated forms of encephalitis.
\newblock In {\em Reference Module in Neuroscience and Biobehavioral
  Psychology}. Elsevier, 2018.

\bibitem{ptavcek2011dopamine}
Radek Pt{\'a}{\v{c}}ek, Hana Ku{\v{z}}elov{\'a}, and George~B Stefano.
\newblock Dopamine d4 receptor gene drd4 and its association with psychiatric
  disorders.
\newblock {\em Medical science monitor: international medical journal of
  experimental and clinical research}, 17(9):RA215, 2011.

\bibitem{di2014dopamine}
Patricia Di~Ciano, David~K Grandy, and Bernard Le~Foll.
\newblock Dopamine d4 receptors in psychostimulant addiction.
\newblock {\em Advances in pharmacology}, 69:301--321, 2014.

\bibitem{hope1998cocaine}
Bruce~T Hope.
\newblock Cocaine and the ap-1 transcription factor complex.
\newblock {\em Annals of the New York Academy of Sciences}, 844(1):1--6, 1998.

\bibitem{robison2011transcriptional}
Alfred~J Robison and Eric~J Nestler.
\newblock Transcriptional and epigenetic mechanisms of addiction.
\newblock {\em Nature reviews neuroscience}, 12(11):623--637, 2011.

\bibitem{schmitt2013nonclassical}
Kyle~C Schmitt, Richard~B Rothman, and Maarten~EA Reith.
\newblock Nonclassical pharmacology of the dopamine transporter: atypical
  inhibitors, allosteric modulators, and partial substrates.
\newblock {\em Journal of Pharmacology and Experimental Therapeutics},
  346(1):2--10, 2013.

\bibitem{reith2015behavioral}
Maarten~EA Reith, Bruce~E Blough, Weimin~C Hong, Kymry~T Jones, Kyle~C Schmitt,
  Michael~H Baumann, John~S Partilla, Richard~B Rothman, and Jonathan~L Katz.
\newblock Behavioral, biological, and chemical perspectives on atypical agents
  targeting the dopamine transporter.
\newblock {\em Drug and alcohol dependence}, 147:1--19, 2015.

\bibitem{jordan2019progress}
Chloe~J Jordan, Jianjing Cao, Amy~Hauck Newman, and Zheng-Xiong Xi.
\newblock Progress in agonist therapy for substance use disorders: Lessons
  learned from methadone and buprenorphine.
\newblock {\em Neuropharmacology}, 158:107609, 2019.

\bibitem{hancox2008herg}
Jules~C Hancox, Mark~J McPate, Aziza El~Harchi, and Yi~hong Zhang.
\newblock The herg potassium channel and herg screening for drug-induced
  torsades de pointes.
\newblock {\em Pharmacology \& therapeutics}, 119(2):118--132, 2008.

\bibitem{cavalluzzi2020human}
Maria~Maddalena Cavalluzzi, Paola Imbrici, Roberta Gualdani, Angela Stefanachi,
  Giuseppe~Felice Mangiatordi, Giovanni Lentini, and Orazio Nicolotti.
\newblock Human ether-{\`a}-go-go-related potassium channel: Exploring sar to
  improve drug design.
\newblock {\em Drug discovery today}, 25(2):344--366, 2020.

\bibitem{food2005international}
Food, HHS Drug~Administration, et~al.
\newblock International conference on harmonisation; guidance on s7b
  nonclinical evaluation of the potential for delayed ventricular
  repolarization (qt interval prolongation) by human pharmaceuticals;
  availability. notice.
\newblock {\em Federal register}, 70(202):61133--61134, 2005.

\bibitem{szklarczyk2019string}
Damian Szklarczyk, Annika~L Gable, David Lyon, Alexander Junge, Stefan Wyder,
  Jaime Huerta-Cepas, Milan Simonovic, Nadezhda~T Doncheva, John~H Morris, Peer
  Bork, et~al.
\newblock String v11: protein--protein association networks with increased
  coverage, supporting functional discovery in genome-wide experimental
  datasets.
\newblock {\em Nucleic acids research}, 47(D1):D607--D613, 2019.

\bibitem{ghislat2021recent}
Ghita Ghislat, Taufiq Rahman, and Pedro~J Ballester.
\newblock Recent progress on the prospective application of machine learning to
  structure-based virtual screening.
\newblock {\em Current Opinion in Chemical Biology}, 65:28--34, 2021.

\bibitem{gao2020repositioning}
Kaifu Gao, Duc~Duy Nguyen, Jiahui Chen, Rui Wang, and Guo-Wei Wei.
\newblock Repositioning of 8565 existing drugs for covid-19.
\newblock {\em The journal of physical chemistry letters}, 11(13):5373--5382,
  2020.

\bibitem{gao2020generative}
Kaifu Gao, Duc~D Nguyen, Meihua Tu, and Guo-Wei Wei.
\newblock Generative network complex for the automated generation of druglike
  molecules.
\newblock {\em Journal of Chemical Information and Modeling},
  60(12):5682--5698, 2020.

\bibitem{daina2017swissadme}
Antoine Daina, Olivier Michielin, and Vincent Zoete.
\newblock Swissadme: a free web tool to evaluate pharmacokinetics,
  drug-likeness and medicinal chemistry friendliness of small molecules.
\newblock {\em Scientific reports}, 7(1):1--13, 2017.

\bibitem{gao20202d}
Kaifu Gao, Duc~Duy Nguyen, Vishnu Sresht, Alan~M. Mathiowetz, Meihua Tu, and
  Guo-Wei Wei.
\newblock Are 2d fingerprints still valuable for drug discovery?
\newblock {\em Phys. Chem. Chem. Phys.}, 22:8373--8390, 2020.

\bibitem{brown2001regulation}
Jeffrey~M Brown, Glen~R Hanson, and Annette~E Fleckenstein.
\newblock Regulation of the vesicular monoamine transporter-2: a novel
  mechanism for cocaine and other psychostimulants.
\newblock {\em Journal of Pharmacology and Experimental Therapeutics},
  296(3):762--767, 2001.

\bibitem{mash2003cocaine}
Deborah~C Mash, Qinjie Ouyang, John Pablo, Margaret Basile, Sari Izenwasser,
  Abraham Lieberman, and Richard~J Perrin.
\newblock Cocaine abusers have an overexpression of $\alpha$-synuclein in
  dopamine neurons.
\newblock {\em Journal of Neuroscience}, 23(7):2564--2571, 2003.

\bibitem{vrana1993chronic}
Sheila~L Vrana, Kent~E Vrana, Timothy~R Koves, James~E Smith, and Steven~I
  Dworkin.
\newblock Chronic cocaine administration increases cns tyrosine hydroxylase
  enzyme activity and mrna levels and tryptophan hydroxylase enzyme activity
  levels.
\newblock {\em Journal of neurochemistry}, 61(6):2262--2268, 1993.

\bibitem{gaulton2012chembl}
Anna Gaulton, Louisa~J Bellis, A~Patricia Bento, Jon Chambers, Mark Davies,
  Anne Hersey, Yvonne Light, Shaun McGlinchey, David Michalovich, Bissan
  Al-Lazikani, et~al.
\newblock Chembl: a large-scale bioactivity database for drug discovery.
\newblock {\em Nucleic acids research}, 40(D1):D1100--D1107, 2012.

\bibitem{smith2014synaptic}
Alexander~CW Smith, Yonatan~M Kupchik, Michael~D Scofield, Cassandra~D Gipson,
  Armina Wiggins, Charles~A Thomas, and Peter~W Kalivas.
\newblock Synaptic plasticity mediating cocaine relapse requires matrix
  metalloproteinases.
\newblock {\em Nature neuroscience}, 17(12):1655--1657, 2014.

\bibitem{efange1998modified}
Simon~MN Efange, Deborah~C Mash, Anil~B Khare, and Quinjie Ouyang.
\newblock Modified ibogaine fragments: Synthesis and preliminary
  pharmacological characterization of 3-ethyl-5-phenyl-1, 2, 3, 4, 5,
  6-hexahydroazepino [4, 5-b] benzothiophenes.
\newblock {\em Journal of medicinal chemistry}, 41(23):4486--4491, 1998.

\bibitem{koenig2015anti}
Xaver Koenig and Karlheinz Hilber.
\newblock The anti-addiction drug ibogaine and the heart: a delicate relation.
\newblock {\em Molecules}, 20(2):2208--2228, 2015.

\bibitem{ballon2006systematic}
Jacob~S Ballon and David Feifel.
\newblock A systematic review of modafinil: potential clinical uses and
  mechanisms of action.
\newblock {\em Journal of clinical Psychiatry}, 67(4):554--566, 2006.

\bibitem{dackis2003modafinil}
Charles~A Dackis, Kevin~G Lynch, Elmer Yu, Frederick~F Samaha, Kyle~M Kampman,
  James~W Cornish, Amy Rowan, Sabrina Poole, Lenae White, and Charles~P
  O'Brien.
\newblock Modafinil and cocaine: a double-blind, placebo-controlled drug
  interaction study.
\newblock {\em Drug and alcohol dependence}, 70(1):29--37, 2003.

\bibitem{zhang2017novel}
Hai-Ying Zhang, Guo-Hua Bi, Hong-Ju Yang, Yi~He, Gilbert Xue, Jiajing Cao,
  Gianluigi Tanda, Eliot~L Gardner, Amy~Hauck Newman, and Zheng-Xiong Xi.
\newblock The novel modafinil analog, jjc8-016, as a potential cocaine abuse
  pharmacotherapeutic.
\newblock {\em Neuropsychopharmacology}, 42(9):1871--1883, 2017.

\bibitem{tunstall2018atypical}
Brendan~J Tunstall, Chelsea~P Ho, Jianjing Cao, Jana{\'\i}na~CM Vendruscolo,
  Brooke~E Schmeichel, Rachel~D Slack, Gianluigi Tanda, Alexandra~J Gadiano,
  Rana Rais, Barbara~S Slusher, et~al.
\newblock Atypical dopamine transporter inhibitors attenuate compulsive-like
  methamphetamine self-administration in rats.
\newblock {\em Neuropharmacology}, 131:96--103, 2018.

\bibitem{newman2019translating}
Amy~Hauck Newman, Jianjing Cao, Jacqueline~D Keighron, Chloe~J Jordan, Guo-Hua
  Bi, Ying Liang, Ara~M Abramyan, Alicia~J Avelar, Christopher~W Tschumi,
  Michael~J Beckstead, et~al.
\newblock Translating the atypical dopamine uptake inhibitor hypothesis toward
  therapeutics for treatment of psychostimulant use disorders.
\newblock {\em Neuropsychopharmacology}, 44(8):1435--1444, 2019.

\bibitem{slack2019structure}
Rachel~D Slack, Therese~C Ku, Jianjing Cao, JoLynn~B Giancola, Alessandro
  Bonifazi, Claus~J Loland, Alexandra Gadiano, Jenny Lam, Rana Rais, Barbara~S
  Slusher, et~al.
\newblock Structure--activity relationships for a series of (bis
  (4-fluorophenyl) methyl) sulfinyl alkyl alicyclic amines at the dopamine
  transporter: Functionalizing the terminal nitrogen affects affinity,
  selectivity, and metabolic stability.
\newblock {\em Journal of medicinal chemistry}, 63(5):2343--2357, 2019.

\bibitem{penetar2006benztropine}
David~M Penetar, Alison~R Looby, Zhaohui Su, Leslie~H Lundahl, Monika
  Er{\"o}s-Sarnyai, Jane~F McNeil, and Scott~E Lukas.
\newblock Benztropine pretreatment does not affect responses to acute cocaine
  administration in human volunteers.
\newblock {\em Human Psychopharmacology: Clinical and Experimental},
  21(8):549--559, 2006.

\bibitem{velazquez2013atypical}
Clara Vel{\'a}zquez-S{\'a}nchez, Jos{\'e}~M Garc{\'\i}a-Verdugo, Juan Murga,
  and Juan~J Canales.
\newblock The atypical dopamine transport inhibitor, jhw 007, prevents
  amphetamine-induced sensitization and synaptic reorganization within the
  nucleus accumbens.
\newblock {\em Progress in Neuro-Psychopharmacology and Biological Psychiatry},
  44:73--80, 2013.

\bibitem{katz2003behavioral}
Jonathan~L Katz, Therissa~A Libby, Theresa Kopajtic, Stephen~M Husbands, and
  Amy~Hauck Newman.
\newblock Behavioral effects of rimcazole analogues alone and in combination
  with cocaine.
\newblock {\em European journal of pharmacology}, 468(2):109--119, 2003.

\bibitem{arunotayanun2013analysis}
Warunya Arunotayanun, Jeffrey~W Dalley, Xi-Ping Huang, Vincent Setola, Ric
  Treble, Leslie Iversen, Bryan~L Roth, and Simon Gibbons.
\newblock An analysis of the synthetic tryptamines amt and 5-meo-dalt: emerging
  ‘novel psychoactive drugs’.
\newblock {\em Bioorganic \& medicinal chemistry letters}, 23(11):3411--3415,
  2013.

\bibitem{grabowski2001dextroamphetamine}
John Grabowski, Howard Rhoades, Joy Schmitz, Angela Stotts, Lee~Ann Daruzska,
  Dan Creson, and F~Gerard Moeller.
\newblock Dextroamphetamine for cocaine-dependence treatment: a double-blind
  randomized clinical trial.
\newblock {\em Journal of clinical psychopharmacology}, 21(5):522--526, 2001.

\bibitem{guerrero2019modulation}
Roc{\'\i}o Guerrero-Bautista, Bruno~Ribeiro Do~Couto, Juana~M Hidalgo,
  Francisco~Jos{\'e} C{\'a}rceles-Moreno, Guillermo Molina, M~Luisa Laorden,
  Cristina N{\'u}{\~n}ez, and M~Victoria Milan{\'e}s.
\newblock Modulation of stress-and cocaine prime-induced reinstatement of
  conditioned place preference after memory extinction through dopamine d3
  receptor.
\newblock {\em Progress in Neuro-Psychopharmacology and Biological Psychiatry},
  92:308--320, 2019.

\bibitem{di2003attenuation}
Patricia Di~Ciano, Rachel~J Underwood, Jim~J Hagan, and Barry~J Everitt.
\newblock Attenuation of cue-controlled cocaine-seeking by a selective d 3
  dopamine receptor antagonist sb-277011-a.
\newblock {\em Neuropsychopharmacology}, 28(2):329--338, 2003.

\bibitem{xi2007pharmacological}
Zheng-Xiong Xi and Eliot~L Gardner.
\newblock Pharmacological actions of ngb 2904, a selective dopamine d3 receptor
  antagonist, in animal models of drug addiction.
\newblock {\em CNS drug reviews}, 13(2):240--259, 2007.

\bibitem{higley2011pg01037}
Amanda~E Higley, Krista Spiller, Peter Grundt, Amy~Hauck Newman, Stephen~W
  Kiefer, Zheng-Xiong Xi, and Eliot~L Gardner.
\newblock Pg01037, a novel dopamine d3 receptor antagonist, inhibits the
  effects of methamphetamine in rats.
\newblock {\em Journal of psychopharmacology}, 25(2):263--273, 2011.

\bibitem{shaik2019investigation}
Anver~Basha Shaik, Vivek Kumar, Alessandro Bonifazi, Adrian~M Guerrero,
  Sophie~L Cemaj, Alexandra Gadiano, Jenny Lam, Zheng-Xiong Xi, Rana Rais,
  Barbara~S Slusher, et~al.
\newblock Investigation of novel primary and secondary pharmacophores and
  3-substitution in the linking chain of a series of highly selective and
  bitopic dopamine d3 receptor antagonists and partial agonists.
\newblock {\em Journal of medicinal chemistry}, 62(20):9061--9077, 2019.

\bibitem{orio2010dopamine}
Laura Orio, Sunmee Wee, Amy~H Newman, Luigi Pulvirenti, and George~F Koob.
\newblock The dopamine d3 receptor partial agonist cjb090 and antagonist
  pg01037 decrease progressive ratio responding for methamphetamine in rats
  with extended-access.
\newblock {\em Addiction biology}, 15(3):312, 2010.

\bibitem{roman2013cariprazine}
V~Roman, I~Gyertyan, K~Saghy, B~Kiss, and ZS~Szombathelyi.
\newblock Cariprazine (rgh-188), a d 3-preferring dopamine d 3/d 2 receptor
  partial agonist antipsychotic candidate demonstrates anti-abuse potential in
  rats.
\newblock {\em Psychopharmacology}, 226(2):285--293, 2013.

\bibitem{wang2015neurotransmitter}
Kevin~H Wang, Aravind Penmatsa, and Eric Gouaux.
\newblock Neurotransmitter and psychostimulant recognition by the dopamine
  transporter.
\newblock {\em Nature}, 521(7552):322--327, 2015.

\bibitem{cervo2003cocaine}
L~Cervo, F~Carnovali, JA~Stark, and T~Mennini.
\newblock Cocaine-seeking behavior in response to drug-associated stimuli in
  rats: involvement of d 3 and d 2 dopamine receptors.
\newblock {\em Neuropsychopharmacology}, 28(6):1150--1159, 2003.

\bibitem{shannon2003cytoscape}
Paul Shannon, Andrew Markiel, Owen Ozier, Nitin~S Baliga, Jonathan~T Wang,
  Daniel Ramage, Nada Amin, Benno Schwikowski, and Trey Ideker.
\newblock Cytoscape: a software environment for integrated models of
  biomolecular interaction networks.
\newblock {\em Genome research}, 13(11):2498--2504, 2003.

\bibitem{cho2014learning}
Kyunghyun Cho, Bart Van~Merri{\"e}nboer, Caglar Gulcehre, Dzmitry Bahdanau,
  Fethi Bougares, Holger Schwenk, and Yoshua Bengio.
\newblock Learning phrase representations using rnn encoder-decoder for
  statistical machine translation.
\newblock {\em arXiv preprint arXiv:1406.1078}, 2014.

\bibitem{hochreiter1997long}
Sepp Hochreiter and J{\"u}rgen Schmidhuber.
\newblock Long short-term memory.
\newblock {\em Neural computation}, 9(8):1735--1780, 1997.

\bibitem{rogers2010extended}
David Rogers and Mathew Hahn.
\newblock Extended-connectivity fingerprints.
\newblock {\em Journal of chemical information and modeling}, 50(5):742--754,
  2010.

\bibitem{hall1995electrotopological}
Lowell~H Hall and Lemont~B Kier.
\newblock Electrotopological state indices for atom types: a novel combination
  of electronic, topological, and valence state information.
\newblock {\em Journal of Chemical Information and Computer Sciences},
  35(6):1039--1045, 1995.

\bibitem{landrum2006rdkit}
Greg Landrum et~al.
\newblock Rdkit: Open-source cheminformatics, 2006.

\bibitem{pedregosa2011scikit}
Fabian Pedregosa, Ga{\"e}l Varoquaux, Alexandre Gramfort, Vincent Michel,
  Bertrand Thirion, Olivier Grisel, Mathieu Blondel, Peter Prettenhofer, Ron
  Weiss, Vincent Dubourg, et~al.
\newblock Scikit-learn: Machine learning in python.
\newblock {\em Journal of machine learning research}, 12(Oct):2825--2830, 2011.

\bibitem{di2006biological}
Li~Di and Edward~H Kerns.
\newblock Biological assay challenges from compound solubility: strategies for
  bioassay optimization.
\newblock {\em Drug discovery today}, 11(9-10):446--451, 2006.

\bibitem{xiong2021admetlab}
Guoli Xiong, Zhenxing Wu, Jiacai Yi, Li~Fu, Zhijiang Yang, Changyu Hsieh,
  Mingzhu Yin, Xiangxiang Zeng, Chengkun Wu, Aiping Lu, et~al.
\newblock Admetlab 2.0: an integrated online platform for accurate and
  comprehensive predictions of admet properties.
\newblock {\em Nucleic Acids Research}, 2021.

\bibitem{fant2019toward}
Kuo~Hao Lee, Andrew~D. Fant, Jiqing Guo, Andy Guan, Joslyn Jung, Mary
  Kudaibergenova, Williams~E. Miranda, Therese Ku, Jianjing Cao, Soren Wacker,
  Henry~J. Duff, Amy~Hauck Newman, Sergei~Y. Noskov, and Lei Shi.
\newblock Toward reducing herg affinities for dat inhibitors with a combined
  machine learning and molecular modeling approach.
\newblock {\em Journal of Chemical Information and Modeling}, 0(0), 2021.

\bibitem{warszycki2021pharm}
Dawid Warszyckia, Łukasz Struskib, Marek Śmiejab, Rafał Kafela, and Rafał
  Kurczaba.
\newblock Pharmacoprint – a combination of pharmacophore fingerprint and
  artificial intelligence as a tool for computeraided drug design.
\newblock 2021.

\bibitem{kalliokoski2013comparability}
Tuomo Kalliokoski, Christian Kramer, Anna Vulpetti, and Peter Gedeck.
\newblock Comparability of mixed ic 50 data--a statistical analysis.
\newblock {\em PloS one}, 8(4):e61007, 2013.

\end{thebibliography}


\begin{thebibliography}{10}

\bibitem{smith2014synaptic}
Alexander~CW Smith, Yonatan~M Kupchik, Michael~D Scofield, Cassandra~D Gipson,
  Armina Wiggins, Charles~A Thomas, and Peter~W Kalivas.
\newblock Synaptic plasticity mediating cocaine relapse requires matrix
  metalloproteinases.
\newblock {\em Nature neuroscience}, 17(12):1655--1657, 2014.

\bibitem{fant2019toward}
Kuo~Hao Lee, Andrew~D. Fant, Jiqing Guo, Andy Guan, Joslyn Jung, Mary
  Kudaibergenova, Williams~E. Miranda, Therese Ku, Jianjing Cao, Soren Wacker,
  Henry~J. Duff, Amy~Hauck Newman, Sergei~Y. Noskov, and Lei Shi.
\newblock Toward reducing herg affinities for dat inhibitors with a combined
  machine learning and molecular modeling approach.
\newblock {\em Journal of Chemical Information and Modeling}, 0(0), 2021.

\bibitem{warszycki2021pharm}
Dawid Warszyckia, Łukasz Struskib, Marek Śmiejab, Rafał Kafela, and Rafał
  Kurczaba.
\newblock Pharmacoprint – a combination of pharmacophore fingerprint and
  artificial intelligence as a tool for computeraided drug design.
\newblock 2021.

\bibitem{sterling2015zinc}
Teague Sterling and John~J Irwin.
\newblock Zinc 15--ligand discovery for everyone.
\newblock {\em Journal of chemical information and modeling},
  55(11):2324--2337, 2015.

\bibitem{kim2016pubchem}
Sunghwan Kim, Paul~A Thiessen, Evan~E Bolton, Jie Chen, Gang Fu, Asta
  Gindulyte, Lianyi Han, Jane He, Siqian He, Benjamin~A Shoemaker, et~al.
\newblock Pubchem substance and compound databases.
\newblock {\em Nucleic acids research}, 44(D1):D1202--D1213, 2016.

\bibitem{gaulton2012chembl}
Anna Gaulton, Louisa~J Bellis, A~Patricia Bento, Jon Chambers, Mark Davies,
  Anne Hersey, Yvonne Light, Shaun McGlinchey, David Michalovich, Bissan
  Al-Lazikani, et~al.
\newblock Chembl: a large-scale bioactivity database for drug discovery.
\newblock {\em Nucleic acids research}, 40(D1):D1100--D1107, 2012.

\bibitem{landrum2006rdkit}
Greg Landrum et~al.
\newblock Rdkit: Open-source cheminformatics, 2006.

\bibitem{hochreiter1997long}
Sepp Hochreiter and J{\"u}rgen Schmidhuber.
\newblock Long short-term memory.
\newblock {\em Neural computation}, 9(8):1735--1780, 1997.

\bibitem{gobbi1998genetic}
Alberto Gobbi and Dieter Poppinger.
\newblock Genetic optimization of combinatorial libraries.
\newblock {\em Biotechnology and bioengineering}, 61(1):47--54, 1998.

\bibitem{gao20202d}
Kaifu Gao, Duc~Duy Nguyen, Vishnu Sresht, Alan~M. Mathiowetz, Meihua Tu, and
  Guo-Wei Wei.
\newblock Are 2d fingerprints still valuable for drug discovery?
\newblock {\em Phys. Chem. Chem. Phys.}, 22:8373--8390, 2020.

\end{thebibliography}

\end{document}


\title{
Supporting information for \\
Proteome-informed machine learning studies of cocaine addiction}

 \author{ Kaifu Gao$^1$, Dong Chen$^1$, Alfred J  Robison$^2$, and Guo-Wei Wei$^{1,3,4}$\footnote{
		Corresponding author.		Email: wei@math.msu.edu} \\
	$^1$ Department of Mathematics, \\
	Michigan State University, MI 48824, USA.\\
	$^2$ Department of Physiology,\\
	Michigan State University, MI 48824, USA.\\
	$^3$Department of Electrical and Computer Engineering,\\
	Michigan State University, MI 48824, USA. \\
	$^4$ Department of Biochemistry and Molecular Biology,\\
	Michigan State University, MI 48824, USA. \\
}

\date{\today} 

\maketitle

\pagenumbering{roman}
\begin{verbatim}
\end{verbatim}

\newpage

{\setcounter{tocdepth}{4} \tableofcontents}
\newpage


\setcounter{page}{1}
\renewcommand{\thepage}{{\arabic{page}}}


\section{Additional cross-target predictions}

\subsection{Additional cross-target BA correlation analysis of binding site similarity }
\begin{figure}[ht!]
	\centering
	\includegraphics[width=0.88\textwidth]{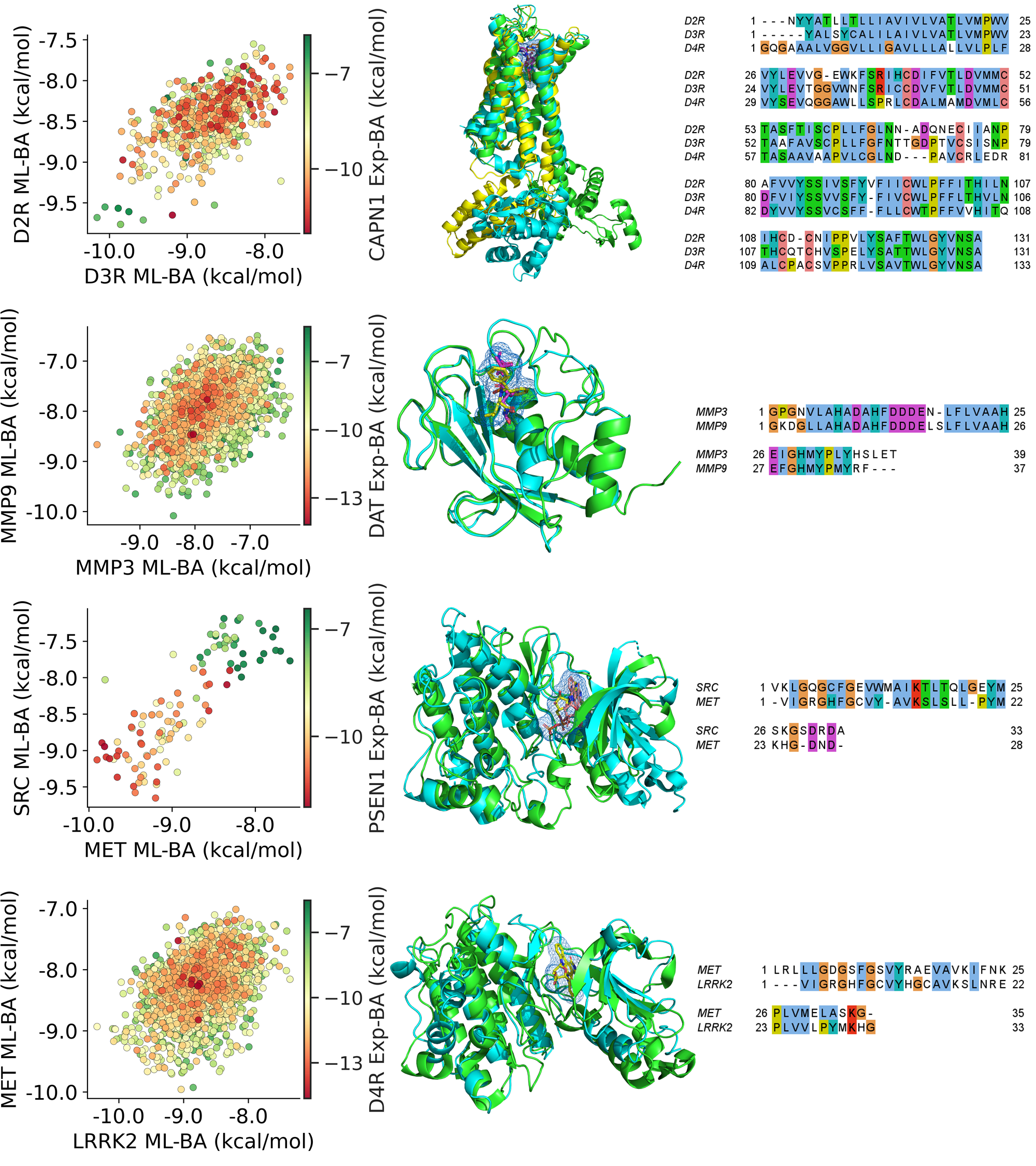}
	\caption{Examples of positive binding affinity (BA) correlation revealing binding site similarity. Each row is for a pair of correlated targets: (1) D$_2$R, D$_3$R, and D$_4$R, (2) MMP3 and MMP9, (3) MET and SPC. (4) LRRK2 and MET. In each row, the first column is the BA correlation plot of one dataset to a pair of target, while ML-BA represents predicted binding affinity from machine-learning (ML) models. The second column is the 3D alignment of target proteins with ligands indicating the location of the binding sites. The third column is the 2D sequence alignment of the binding sites of the target proteins. The PDB IDs of the protein structures are 6CM4, 3PBL, 5WIV, 1B3D, 1GKC, 1R0P, 1KSW, and 6VP7 for D$_2$R, D$_3$R, D$_4$R, MMP3, MMP9, MET, SPC, and LRRK2, respectively. }
	\label{fig:correlation-alignment}
\end{figure}

We present additional examples of positive cross-target BA correlation revealing the binding-site similarity  in Figure \ref{fig:correlation-alignment}. ML-BA represents machine-learning (ML) predicted BA. The first row depicts the ML-BAs of the compounds in the CAPN1 dataset for the D$_2$R and D$_3$R. The correlation plot and the R of 0.56 indicate the BAs for D$_2$R and D$_3$R are positively correlated. This makes sense because D$_2$R and D$_3$R belong to the dopamine receptor family and have a sequence identity as high as 69.7\%. Additionally, another member of the dopamine receptor family, D$_4$R, also has a binding-site sequence identity of 49.6\% to D$_3$R. As a result, the BAs to D$_4$R are also correlated to that for D$_2$R and D$_3$R (not shown here).

The second row of Figure \ref{fig:correlation-alignment} shows the ML-BAs of the compounds in the DAT dataset for the target MMP3 and MMP9. MMP3 and MMP9 belong to the protein family of matrix metalloproteinases, which play a role in cocaine relapse \cite{smith2014synaptic}. Since they are in the same protein family, we expected to see a clear BA correlation. Indeed, the first graph in the second row reveals BA correlation between these two targets (R=0.48). The second and third graphs are the 3D alignment of their binding domains and 2D sequence alignment of their binding sites, respectively, which suggest their binding sites are highly similar, with a binding-site sequence identity as high as 64.9\%.

In addition to the targets in the same protein family that lead to correlated BAs, we also found some seemingly unrelated targets to show  correlated BAs, indicating these proteins have similar binding domains and binding sites. The third and fourth rows of Figure \ref{fig:correlation-alignment} are two examples in this category. The third row illustrates the BA correlation for targets MET and SRC, while the fourth row is for targets MET and LRRK2. Although MET, SRC, and LRRK2 are structurally distinct, their biological functions are similar in that they are all protein kinases. MET (tyrosine-protein kinase Met) is the tyrosine kinase that transduces signals from the extracellular matrix into the cytoplasm by binding to hepatocyte growth factor ligand. SRC (Proto-oncogene tyrosine-protein kinase Src) is the tyrosine kinase, which is activated following the engagement of many different classes of cellular receptors. In contrast, LRRK2 ( Leucine-rich repeat serine/threonine-protein kinase 2) is the serine/threonine-protein kinase, which phosphorylates a broad range of proteins in multiple processes such as neuronal plasticity, autophagy, and vesicle trafficking. Therefore, as shown in the third and fourth rows of Figure \ref{fig:correlation-alignment}, the binding domains of MET, SRC, and LRRK2 have highly similar 3D conformations, and the 2D sequence identities of SRC and LRRK2 to MET are 50.1\% and 40.2\%, respectively. Thus the R between the BAs to MET and SRC and the R between the BAs for MET and LRRK2 are 0.82 and 0.37, respectively.

\subsection{Additional cross-target BA correlation analysis of  side effects and repurposing potential}

\begin{figure}[ht!]
	\centering
	\includegraphics[width=0.90\textwidth]{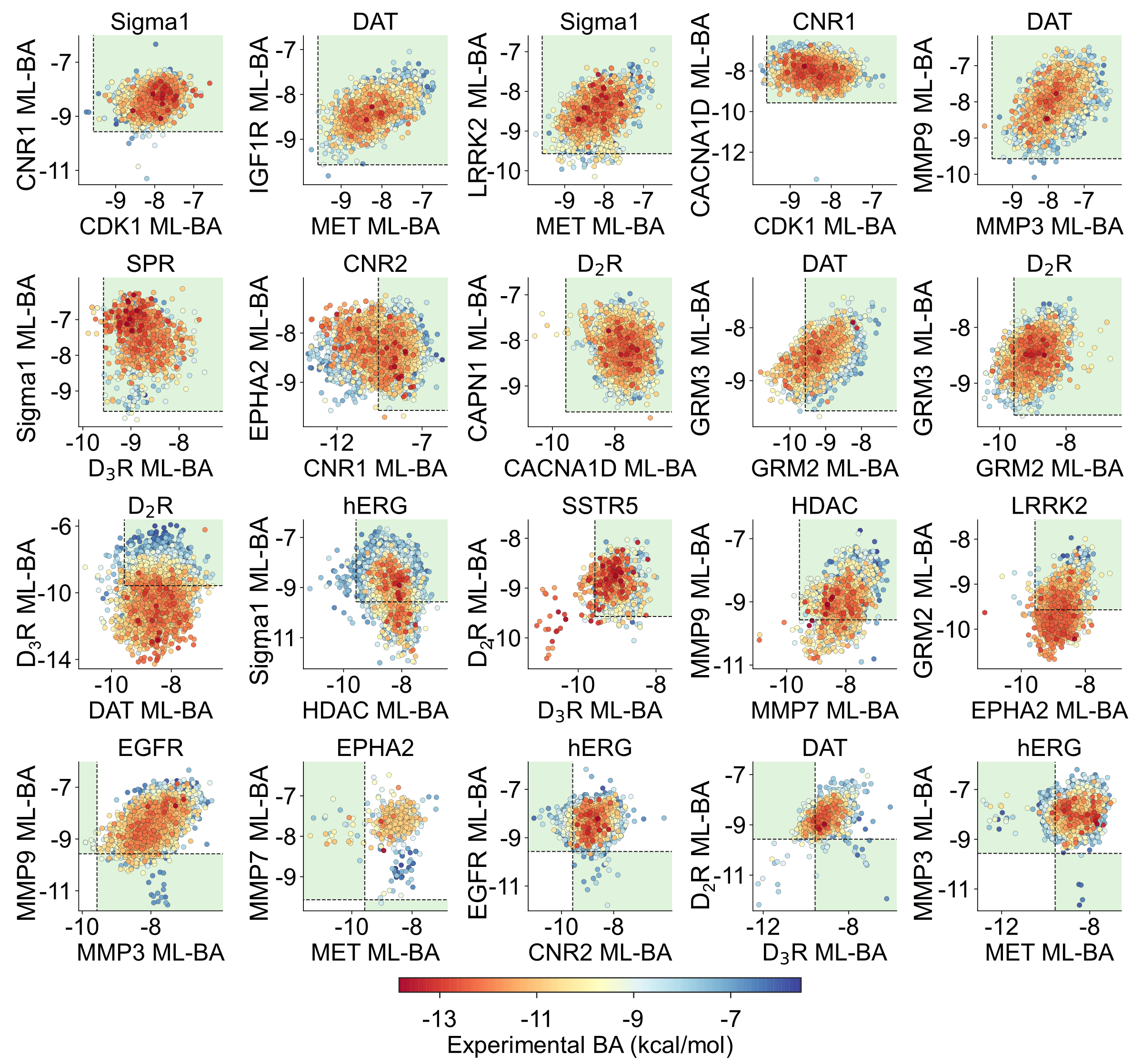}
	\caption{Examples of cross-target BA predictions to detect side effects or repurposing potential. In each plot, the title is the designated target of the dataset. The colors of points represent the experimental BAs to the designated target. The $x-$ and $y-$axes indicate the ML-BA to two other proteins. The first row exhibits the examples without strong side effects to either of the two other proteins. The second row shows the situations of strong interactions with one protein. The third row represents the examples with strong side effects to both of the two proteins. The green frames in these three rows outline  the optimal ranges without side effects to both the two targets ($x$ $>$ -9.57 kcal/mol and $y$ $>$ -9.57 kcal/mol). In the plots of the fourth row, some inhibitors are not so potent to their designated targets but predicted to strongly interact with another protein, which suggests drug repurposing potential. The two green frames in each subplot of the fourth row outline the BA domains with repurposing potential, which represent compounds that have the potency to one target (BA values $<$ -9.57 kcal/mol) and do not show strong side effects to the other target (BA values $>$ -9.57 kcal/mol).}
	\label{fig:side-effect-cross-predict}
\end{figure}

Figure \ref{fig:side-effect-cross-predict} depicts more examples of our cross-target BA predictions detecting side effects and repurposing potential. In each of our studies, three targets are involved: its designated target and two other targets.

The first row reflects the situation for inhibitors without strong side effects to either of two other targets. In the five plots, all potent inhibitors to their designated targets, which are represented by red or even deep red points, are predicted to have much weaker ($>$ -9.0 kcal/mol) affinities to the other two proteins. Therefore, in these cases, we anticipated that these potent inhibitors would not have strong side effects through other targets.

The second row contains examples with potential side effects on one protein. For instance, the second plot of this row shows that potent inhibitors of protein CNR2 are unlikely to bind to EPHA2. However, some potent inhibitors have strong predicted BAs ($\approx$ -12 kcal/mol) to CNR1. This side effect is expected, as CNR1 and CNR2 are similar and both cannabinoid receptors.

The third row is the worst case, in which side effects happen to both of the two other proteins. The typical examples are the third and fourth plots, which suggest the side effects to D$_2$R and D$_3$R occur simultaneously, so as MMP7 and MMP9. This is the case because D$_2$R and D$_3$R as well as MMP7 and MMP9 are two pairs of similar targets.  

Besides side-effect anticipation, our cross-target BA predictions can also suggest drug repurposing potential. Namely, some inhibitors are not so potent to their designated targets but predicted to strongly interact with another protein. The fourth row provides some examples in this category. In each plot of the fourth row, some weak inhibitors to its designated targets are predicted to potently inhibit another protein. For instance, in the second plot, some weak EPHA2 inhibitors were predicted to bind strongly to MET.

\subsection{BA predictions of targets DAT and hERG}

\begin{figure}[ht!]
	\centering
	\includegraphics[width=0.9\textwidth]{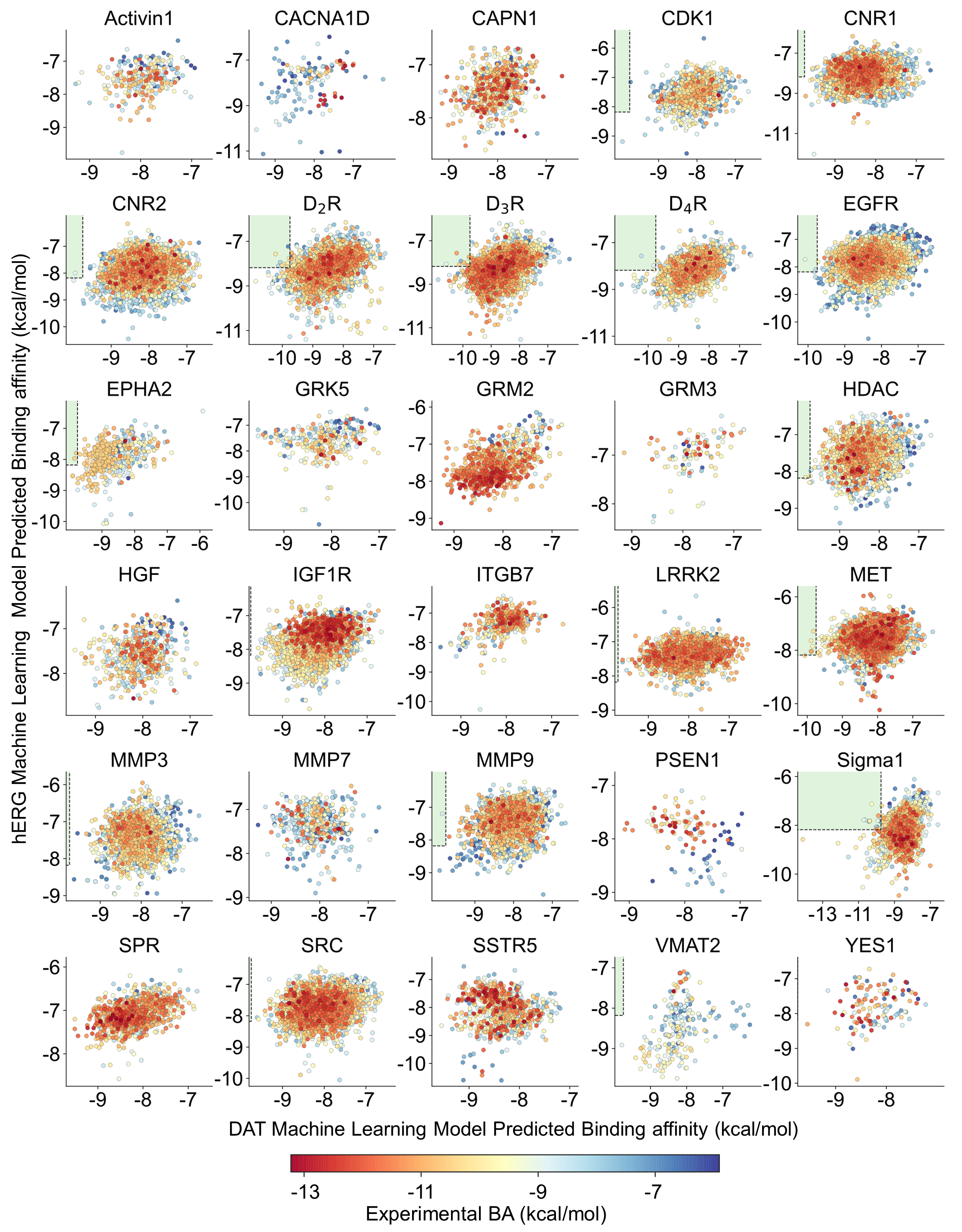}
	\caption{The BA prediction of the compounds from the 30 datasets to DAT and hERG (The datasets for DAT and hERG were excluded). The colors of points represent the experimental binding affinities to their designated targets. The $x-$ and $y-$axes represent their predicted BAs to targets DAT and hERG. The green frames represent the optimal domains with high predicted DAT potency (BA values $<$ -9.57 kcal/mol) and low predicted hERG side effect (BA values $>$ -8.18 kcal/mol).}
	\label{fig:DAT-hERG}
\end{figure}

The BA prediction of the compounds from the 30 datasets to DAT and hERG are shown in Figure \ref{fig:DAT-hERG}.

\subsection{Side-effect  analysis of existing experimental medications for treating cocaine addiction}

\begin{figure}[ht!]
	\centering
	\includegraphics[width=0.9\textwidth]{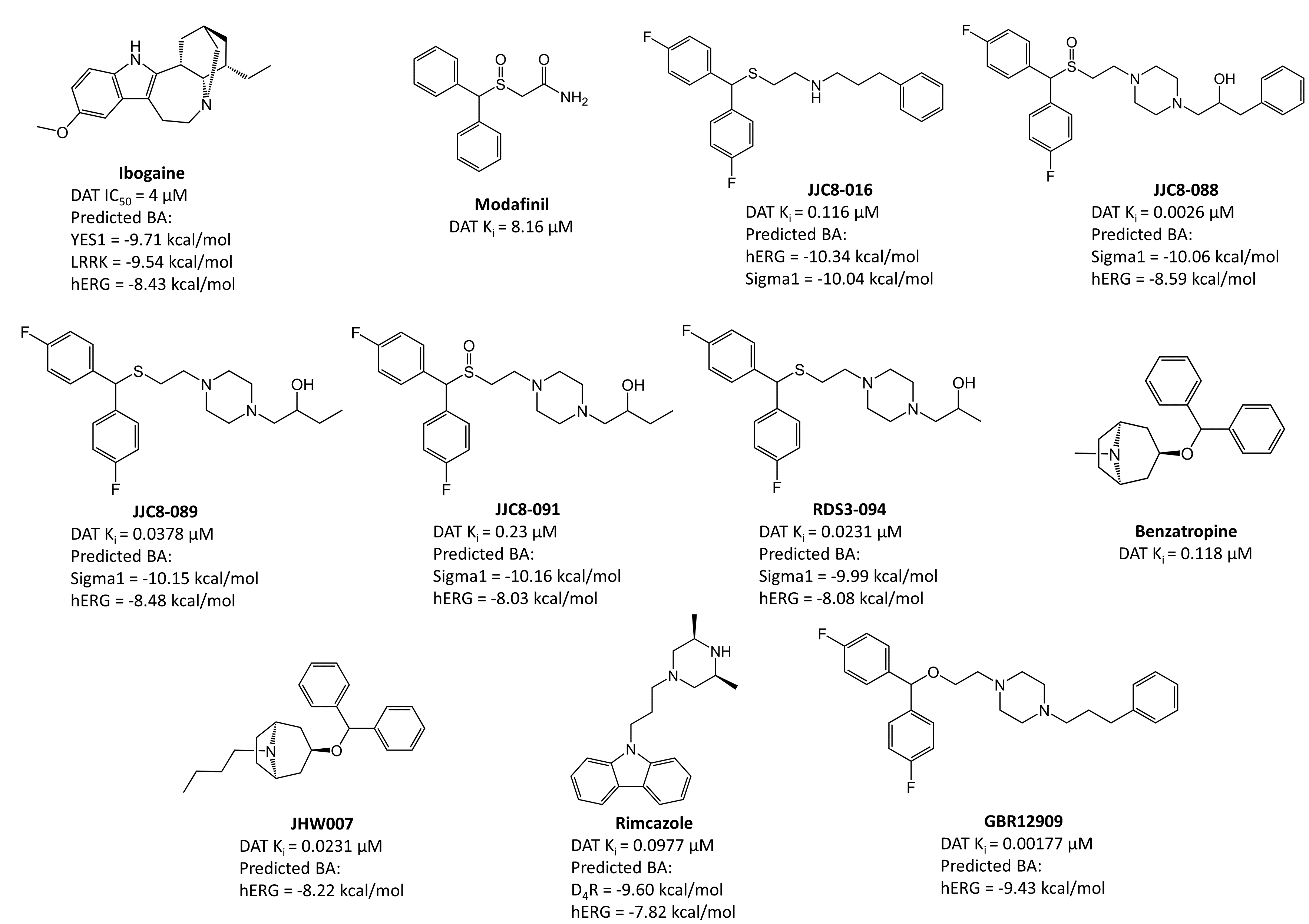}
	\caption{Atypical DAT inhibitors under investigation and their side effect predictions. Only serious side effects with BA < -9.54 kcal/mol and the hERG side effect are shown in the figure.}
	\label{fig:dat-drugs-side_effect}
\end{figure}

\begin{figure}[ht!]
	\centering
	\includegraphics[width=1.0\textwidth]{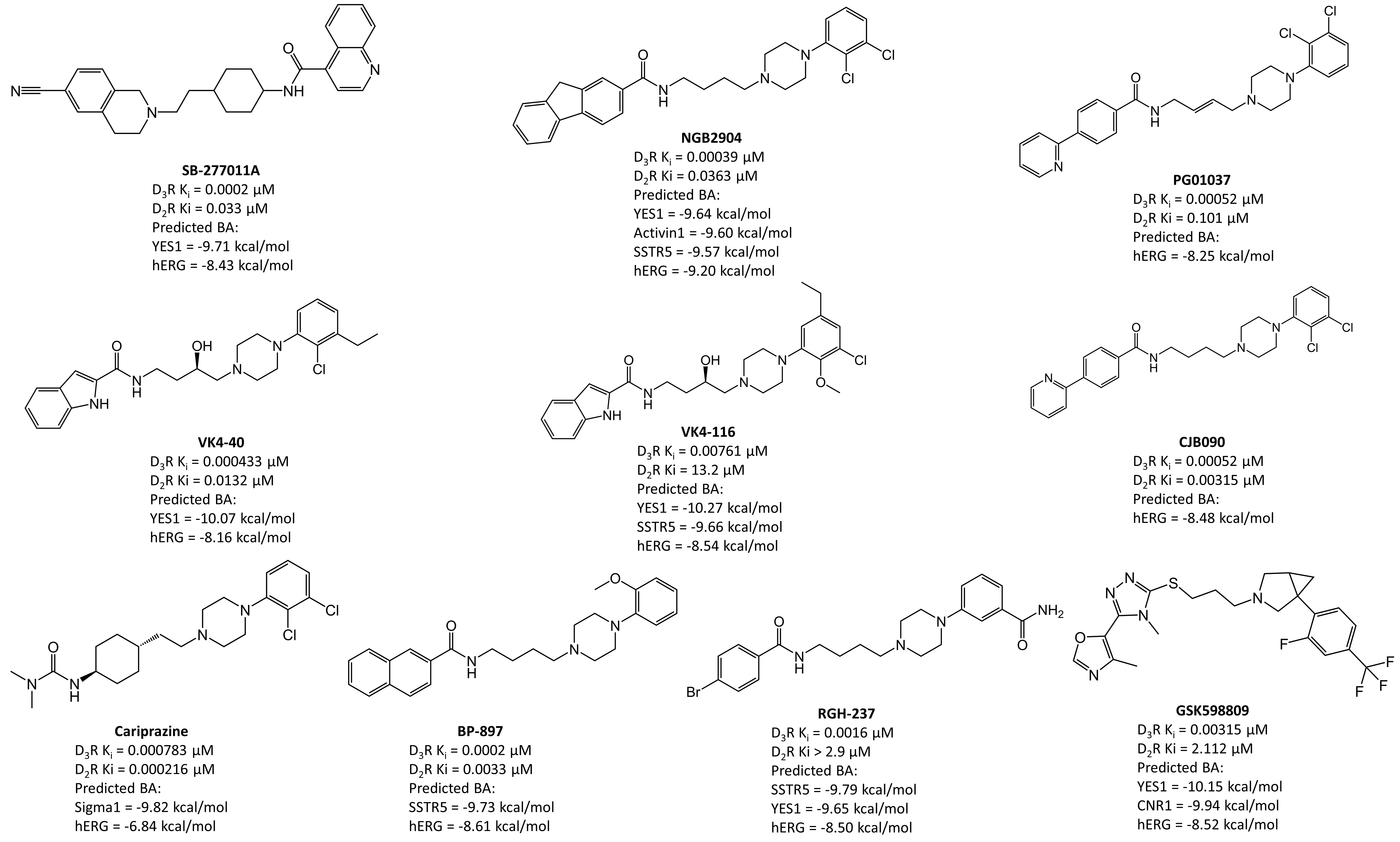}
	\caption{D$_3$R antagonists/partial agonists under investigation and their side effect predictions. Only serious side effects with BA < -9.54 kcal/mol and the hERG side effect are shown in the figure.}
	\label{fig:d3r-drugs-side_effect}
\end{figure}

Figures \ref{fig:dat-drugs-side_effect} and \ref{fig:d3r-drugs-side_effect} show the side-effect prediction to existing experimental medications treating cocaine addiction, which include atypical DAT inhibitors and D$_3$R antagonists/partial agonists.

\subsection{Additional nearly optimal leads obtained from our ML-based screening and  repurposing}

\begin{figure}[ht!]
	\centering
	\includegraphics[width=0.7\textwidth]{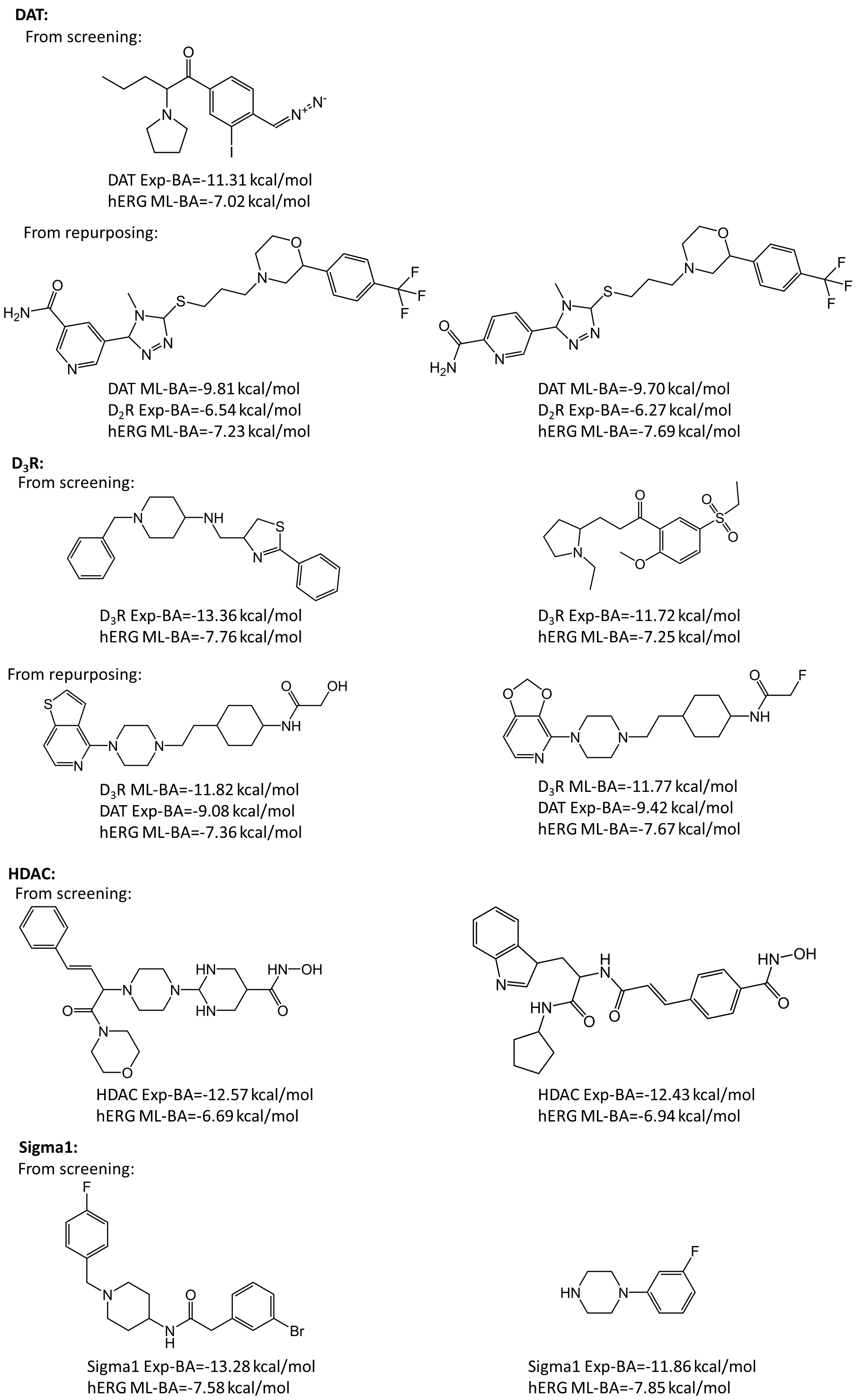}
	\caption{More nearly optimal lead compounds from our ML-based screening or repurposing targeting DAT, D$_3$R, HDAC, and Sigma1.}
	\label{fig:top-compounds}
\end{figure}

More nearly optimal lead compounds from our ML-based screening or repurposing targeting DAT, D$_3$R, HDAC, and Sigma1 are shown in Figure \ref{fig:top-compounds}.

\section{Datasets and performance summary}

Some of cocaine-related datasets used in this work  were adopted from the literature \cite{fant2019toward,warszycki2021pharm} and most cocaine-related datasets were collected from ChEMDL database (https://www.ebi.ac.uk/chembl/). These datasets  are summarized in Table \ref{tab:dateset-sum} and are available  in our 2D Data Library https://weilab.math.msu.edu/DataLibrary/2D/. 

In Table \ref{tab:dateset-sum}, our prediction performances on cocaine-related  datasets are also summarized as Rs of 10-fold cross-validation (R of 10-fold CV).

\begin{table}
	\centering
	\begin{tabular}{c|c|c|c|c}		
		\hline
		Dataset & Target ChEMBLID  & Dataset size  & Binding affinity range (kcal/mol) & R of 10-fold CV   \\ 	
		\hline
		
		DAT binding & CHEMBL238 & 1189 & [-14.18, -2.90] & 0.86 \\
		DAT uptake & CHEMBL238 & 350 & [-12.52, -5.33] & 0.84 \\
		Extended DAT & CHEMBL238 & 2877 & [-14.18, -2.90] & 0.84 \\
		D$_3$R & CHEMBL234 & 4685 & [-14.59, -5.64] & 0.85 \\
		
		D$_2$R from Ref. \cite{warszycki2021pharm} & CHEMBL217 & 3721 & active or inactive & N/A\\
		Extended D$_2$R & CHEMBL217 & 6923 & [-14.41, -5.45] & 0.83\\
		
		D$_4$R & CHEMBL219 & 2411 & [-14.64, -5.52] & 0.78 \\
		HDAC & CHEMBL2093865 & 1925 & [-14.04, -5.86] & 0.85 \\
		Sigma1 & CHEMBL287 & 2388 & [-14.23, -5.90] & 0.80 \\
		Activin receptor 1 & CHEMBL5903 & 257 & [-12.57, -5.45] & 0.74 \\
		VMAT2 & CHEMBL4828 & 248 & [-13.21, -5.73] & 0.71 \\
		CDK1 & CHEMBL308 & 1253 & [-13.36, -5.86] & 0.77\\
		
		CACNA1D & CHEMBL2095229 & 137 & [-15.40, -5.86] & 0.90 \\
		CAPN1 & CHEMBL3891 & 639 & [-12.02, -5.79] & 0.75 \\
		CNR1 & CHEMBL218 & 3922 & [-14.70, -5.45] & 0.85 \\
		CNR2 & CHEMBL253 & 4336 & [-14.79, -5.48] & 0.83 \\
		EGFR & CHEMBL203 & 6693 & [-15.41, -5.67] & 0.86 \\
		EPHA2 & CHEMBL2068 & 490 & [-12.98, -5.86] & 0.80 \\
		GRM2 & CHEMBL5137 & 748 & [-12.62, -5.56] & 0.87 \\
		GRM3 & CHEMBL2888 & 114 & [-12.72, -5.83] & 0.81 \\
		HGF & CHEMBL3717 & 529 & [-13.63, -5.59] & 0.84 \\
		
		IGF1R & CHEMBL1957 & 2450 & [-14.45, -5.45] & 0.92 \\
		ITGB7 & CHEMBL2095184 & 416 & [-12.84, -5.86] & 0.78 \\
		LRRK2 & CHEMBL1075104 & 1871 & [-13.78, -5.45] & 0.85 \\
		MET & CHEMBL3717 & 3347 & [-13.89, -5.59] & 0.86 \\
		
		MMP3 & CHEMBL283 & 1909 & [-14.61, -5.45] & 0.82\\
		MMP7 & CHEMBL4073 & 482 & [-13.55, -5.48] & 0.81 \\
		MMP9 & CHEMBL321 & 2523 & [-15.40, -5.45] & 0.86 \\
		PSEN1 & CHEMBL2473 & 117 & [-11.97, -6.43] & 0.93 \\
		SPR & CHEMBL3988583 & 1026 & [-12.68, -6.41] & 0.76 \\
		SRC & CHEMBL267 & 3268 & [-13.93, -3.54] & 0.86 \\
		SSTR5 & CHEMBL1792 & 788 & [-13.89, -6.09] & 0.86 \\
		YES1 & CHEMBL2073 & 121 & [-13.22, -5.90] & 0.83\\
		GRK5 & CHEMBL5678 & 262 & [-10.36, -5.45] & 0.69 \\
		
		hERG & CHEMBL240 & 2043 & [-12.83, -3.27] &0.83 \\
		Extended hERG & CHEMBL240 & 6298 & [-13.84, -3.27] & 0.80\\
		\hline
	\end{tabular}
	\caption{The summary of datasets investigated in this study and our prediction performance on these datasets.}
	\label{tab:dateset-sum}
\end{table}

\section{Methods and validation}

\subsection{Seq2seq autoencoder model and  latent-vector fingerprints (LV-FPs)}

\begin{figure}[ht!]
	\centering
	\includegraphics[width=0.8\textwidth]{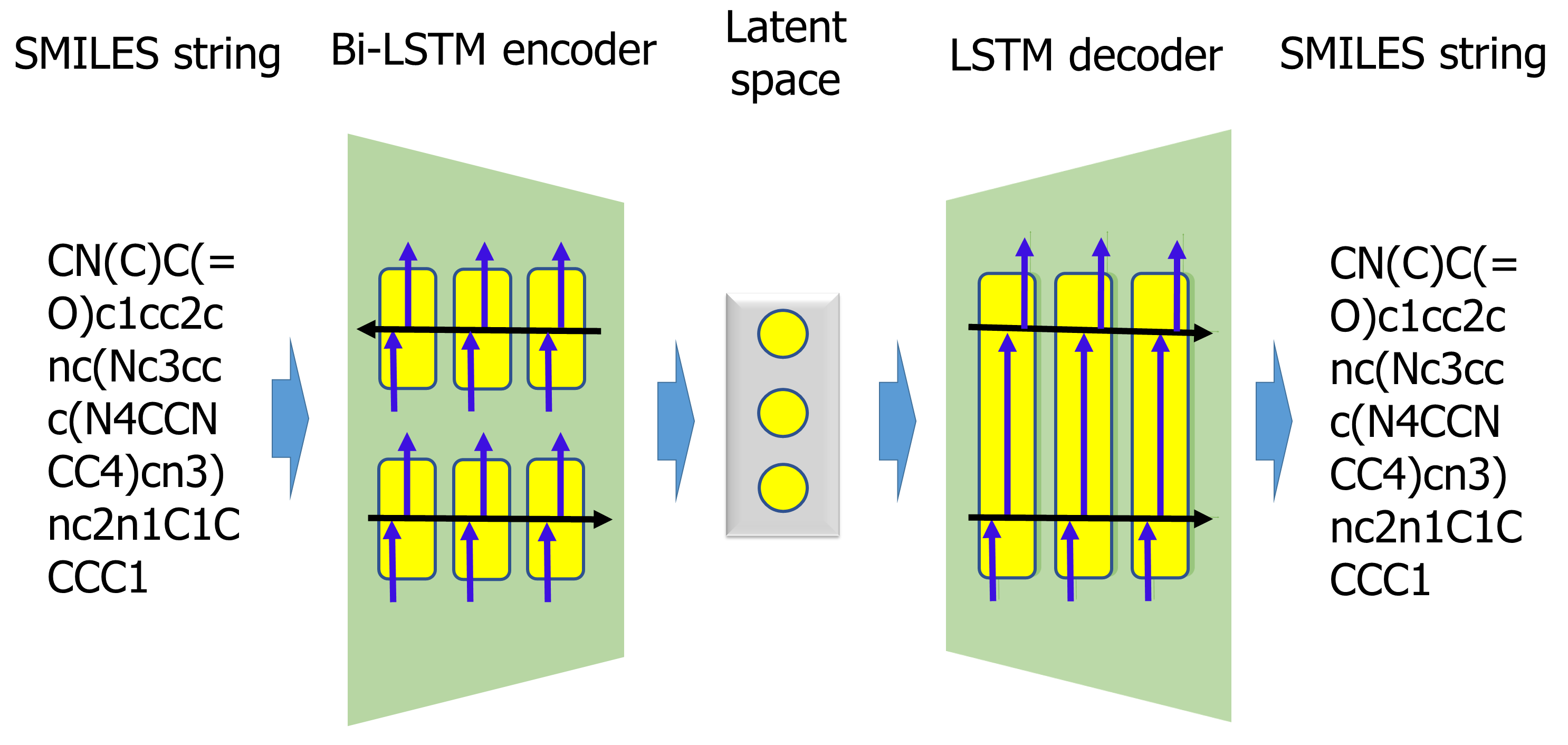}
	\caption{An illustration of our seq2seq autoencoder model for LV-FP generation.}
	\label{fig:seq2seq}
\end{figure}

Our seq2seq autoencoder model for LV-FP generation is illustrated in Figure \ref{fig:seq2seq}. We used almost 104 million SMILES strings to train our seq2seq model. These training compounds are selected from the Zinc15 \cite{sterling2015zinc}, PubChem \cite{kim2016pubchem}, and Chembl27\cite{gaulton2012chembl} databases with the criteria of an RDKit \cite{landrum2006rdkit} predicted log P between -7 to 5, molecular weight between 12 to 600, as well as the length of a canonical smiles string length $<$ 129. Our dictionary considers 61 different symbols from these smiles strings. Our encoder utilizes a bidirectional LSTM based encoder \cite{hochreiter1997long} and an LSTM-based decoder. The dimension of the latent space is 512. The latent space vectors are used to represent molecules. The source code of our autoencoder model is available in https://github.com/KfGao2021/antoencoder-v01.

\subsection{Our predictive power on the benchmark DAT datasets from a previous work}

Lee et al.'s recent work \cite{fant2019toward} provides some benchmark DAT datasets. They classified the DAT data based on their different assay types, namely ``binding" or ``uptake" (see Table \ref{tab:dateset-sum}).

For splitting training and test sets, we applied the same splitting as that in Lee et al.'s work. For each dataset, a random split was performed so that 85\% of the data are used as the training dataset to construct a model, and the remaining 15\% of the data are used as the test set. Since the split is stochastic, we repeated splitting, training, and testing 35 times to eliminate uncertainty following the suggestion by Lee et al.

We tested these datasets by both regression and classification tasks. Since in the classification tasks, the data points with $\rm pK_{i}$ or $\rm pIC_{50}$ between 5 and 6 were erased to define a clearer boundary between
binders and non-binders. The datasets for classification are generally smaller than those for regression.

\subsubsection{Regression tasks}

\begin{table}
	\centering
	\begin{tabular}{c|c|c|c|c|c|c}		
		\hline
		
		Model & Metrics & Fingerprint & Dataset  & Average & Standard deviation & Best   \\ 	
		\hline
		
		\multirow{28}*{\rotatebox{90}{Regression}} & \multirow{14}*{$\rm R^2$}  & \multirow{2}*{LV-FP} & DAT binding & 0.73 & 0.04 & 0.82  \\
		\cline{4-7}
		&&&DAT uptake  & 0.67 & 0.09  & 0.83 \\
		\cline{3-7}
		&&\multirow{2}*{ECFP} & DAT binding & 0.71 & 0.05 & 0.82  \\
		\cline{4-7}
		&&&DAT uptake  & 0.70 & 0.08  & 0.85 \\
		\cline{3-7}
		&&\multirow{2}*{Estate1} & DAT binding & 0.65 & 0.05 & 0.81  \\
		\cline{4-7}
		&&&DAT uptake  & 0.64 & 0.08  & 0.80 \\
		\cline{3-7}
		&&\multirow{2}*{Estate2} & DAT binding & 0.66 & 0.04 & 0.76  \\
		\cline{4-7}
		&&&DAT uptake  & 0.61 & 0.09  & 0.75 \\
		\cline{3-7}
		&&consensus of & DAT binding & 0.74 & 0.05 & 0.84  \\
		\cline{4-7}
		&&LV-FP + ECFP &DAT uptake  & 0.71 & 0.08  & 0.86 \\
		\cline{3-7}
		&&consensus of & DAT binding & 0.73 & 0.04 & 0.82  \\
		\cline{4-7}
		&&all 4 FPs&DAT uptake  & 0.70 & 0.08  & 0.83 \\
		\cline{3-7}
		&&\multirow{2}*{Ref. \cite{fant2019toward}} & DAT binding & 0.71 & 0.04 & 0.80  \\
		\cline{4-7}
		&&&DAT uptake  & 0.66 & 0.11  & -- \\
		\cline{2-7}
		
		& \multirow{14}*{$\rm RMSE$}  & \multirow{2}*{LV-FP} & DAT binding & 0.89 & 0.07 & 0.78  \\
		\cline{4-7}
		&&&DAT uptake  & 0.80 & 0.10  & 0.59 \\
		\cline{3-7}
		&&\multirow{2}*{ECFP} & DAT binding & 0.90 & 0.07 & 0.72  \\
		\cline{4-7}
		&&&DAT uptake  & 0.78 & 0.12  & 0.53 \\
		\cline{3-7}
		&&\multirow{2}*{Estate1} & DAT binding & 0.99 & 0.07 & 0.89  \\
		\cline{4-7}
		&&&DAT uptake  & 0.86 & 0.11  & 0.67 \\
		\cline{3-7}
		&&\multirow{2}*{Estate2} & DAT binding & 0.99 & 0.07 & 0.83  \\
		\cline{4-7}
		&&&DAT uptake  & 0.89 & 0.10  & 0.71 \\
		\cline{3-7}
		&&consensus of & DAT binding & 0.87 & 0.07 & 0.71  \\
		\cline{4-7}
		&&LV-FP + ECFP &DAT uptake  & 0.76 & 0.11  & 0.53 \\
		\cline{3-7}
		&&consensus of & DAT binding &0.89 & 0.07 & 0.74  \\
		\cline{4-7}
		&&all 4 FPs&DAT uptake  & 0.78 & 0.10  & 0.57 \\
		\cline{3-7}
		&&\multirow{2}*{Ref. \cite{fant2019toward}} & DAT binding & 0.89 & 0.05 & --  \\
		\cline{4-7}
		&&&DAT uptake  & 0.82 & 0.11  & -- \\
		\cline{3-7}	
		\hline		
	\end{tabular}
	\caption{The regression results on the benchmark DAT datasets. The unit of RMSE is kcal/mol.}
	\label{tab:dat-benchmark-regress}
\end{table}

The regression tasks adopted the entire sets, so the data sizes for DAT binding and uptake sets are 1138 and 350, respectively. Table \ref{tab:dat-benchmark-regress} represents the regression results from testing these two sets.

Since the ``binding'' set is much larger than the ``uptake'' set, it provides a much large training set,leading to better performance with all fingerprints. The comparison also reveals that, although 35 tests were repeated, the smaller uptake set leads to larger standard deviations of $\rm R^2$ and $\rm RMSE$, which suggests that larger datasets produce more robust and stable results. 

Between different fingerprints, LV-FP derived from our seq2seq model outperforms  traditional 2D fingerprints. For example, on the binding and uptake sets, LV-FP yields $\rm R^2$s 0.73 and 0.67, respectively. By contrast, Lee et al.'s $\rm R^2$s are 0.71 and 0.66, respectively \cite{fant2019toward}. 

We also test three traditional 2D fingerprints (ECFP, Estate1, and Estate2) in our experiment. On both larger (binding) and smaller (uptake) sets, ECFP performs almost  as good as or better than the descriptors in Lee et al.'s work \cite{fant2019toward}. Considering their descriptors integrate  200 fingerprints from Rdkit as well as the Gobbi 2D pharmacophore fingerprint \cite{gobbi1998genetic}, ECFP is quite promising for  predicting DAT inhibitors.

Notably, our study also reveals, although traditional 2D fingerprints always lead to worse prediction power than LV-FP, the consensus with 2D fingerprints could still improve the performance of LV-FP. For instance, on the binding dataset, the consensus with ECFP (consensus of top 2 in Table \ref{tab:dat-benchmark-regress}) could further lift $\rm R^2$ and $RMSE$ from 0.73 and 0.89 to 0.74 and 0.87, respectively. Moreover, the results from Estate1 and Estate2 seem to be much worse. However, the consensus also including these two fingerprints does not ruin the performance of ECFP at all. Therefore, the consensus of different fingerprints is a robust way to enhance the prediction power, which is consistent with our early findings \cite{gao20202d}. 


\subsubsection{Classification tasks}

\begin{table}
	\centering
	\begin{tabular}{c|c|c|c|c|c|c}		
		\hline
		
		Model & Metrics & Fingerprint & Dataset  & Average & Standard deviation & Best   \\ 	
		\hline
		
		\multirow{28}*{\rotatebox{90}{Classification}} & \multirow{14}*{Accuracy}  & \multirow{2}*{LV-FP} & DAT binding & 0.97 & 0.01 & 0.99  \\
		\cline{4-7}
		&&&DAT uptake  & 0.96 & 0.03  & 1.00 \\
		\cline{3-7}
		&&\multirow{2}*{ECFP} & DAT binding & 0.97 & 0.01 & 0.98  \\
		\cline{4-7}
		&&&DAT uptake  & 0.96 & 0.03  & 1.00 \\
		\cline{3-7}
		&&\multirow{2}*{Estate1} & DAT binding & 0.97 & 0.01 & 0.98  \\
		\cline{4-7}
		&&&DAT uptake  & 0.97 & 0.02  & 1.00 \\
		\cline{3-7}
		&&\multirow{2}*{Estate2} & DAT binding & 0.96 & 0.01 & 0.99  \\
		\cline{4-7}
		&&&DAT uptake  & 0.97 & 0.02  & 1.00 \\
		\cline{3-7}
		&&consensus of & DAT binding & 0.97 & 0.01 & 0.98  \\
		\cline{4-7}
		&&LV-FP + ECFP&DAT uptake  & 0.97 & 0.03  & 1.00 \\
		\cline{3-7}
		&&consensus of  & DAT binding & 0.96 & 0.01 & 0.98  \\
		\cline{4-7}
		&&all 4 FPs &DAT uptake  & 0.97 & 0.01  & 1.00 \\
		\cline{3-7}
		&&\multirow{2}*{Ref\cite{fant2019toward}} & DAT binding & 0.96 & 0.02 & --  \\
		\cline{4-7}
		&&&DAT uptake  & 0.95 & 0.03  & -- \\
		\cline{2-7}
		
		& \multirow{14}*{F score}  & \multirow{2}*{LV-FP} & DAT binding & 0.98 & 0.01 & 0.99  \\
		\cline{4-7}
		&&&DAT uptake  & 0.98 & 0.01  & 1.00 \\
		\cline{3-7}
		&&\multirow{2}*{ECFP} & DAT binding & 0.98 & 0.01 & 0.99  \\
		\cline{4-7}
		&&&DAT uptake  & 0.98 & 0.02  & 1.00 \\
		\cline{3-7}
		&&\multirow{2}*{Estate1} & DAT binding & 0.98 & 0.01 & 0.99  \\
		\cline{4-7}
		&&&DAT uptake  & 0.98 & 0.01  & 1.00 \\
		\cline{3-7}
		&&\multirow{2}*{Estate2} & DAT binding & 0.98 & 0.01 & 0.99  \\
		\cline{4-7}
		&&&DAT uptake  & 0.98 & 0.01  & 1.00 \\
		\cline{3-7}
		&&consensus of & DAT binding & 0.98 & 0.01 & 0.99  \\
		\cline{4-7}
		&&LV-FP + ECFP &DAT uptake  & 0.98 & 0.01  & 1.00 \\
		\cline{3-7}
		&&consensus of & DAT binding &0.99 & 0.01 & 1.00  \\
		\cline{4-7}
		&&the 4 FPs&DAT uptake  & 0.99 & 0.01  & 1.00 \\
		\cline{3-7}
		&&\multirow{2}*{Ref\cite{fant2019toward}} & DAT binding & 0.98 & 0.01 & --  \\
		\cline{4-7}
		&&&DAT uptake  & 0.97 & 0.02  & -- \\
		\cline{3-7}
		
		\hline
		
	\end{tabular}
	\caption{Classification results on the benchmark DAT datasets.}
	\label{tab:dat-benchmark-class}
\end{table}

For the classification task, since the data points with $\rm pK_{i}$ or $\rm pIC_{50}$ between 5 and 6 were removed, the data sizes for DAT ``binding'' and ``uptake'' sets are reduced to 887 and 219, respectively. Table \ref{tab:dat-benchmark-class} shows the classification results on these two datasets.

Similar to the results of regression tasks, the results from LV-FP are also better than those from traditional 2D fingerprints. For example, on the ``binding'' and ``uptake'' sets, LV-FP's accuracies are 0.97 and 0.96, respectively. The accuracies  reported by Lee et al. are 0.96 and 0.95 \cite{fant2019toward}. 

We also tested the three traditional 2D fingerprints (ECFP, Estate1, and Estate2) in our experiments. On both larger (binding) and smaller (uptake) datasets, all the three fingerprints have better or the same performance with the descriptors used by Lee et al. 

Additionally, the consensus of LV-FP with 2D fingerprints also brings enhancement to results. For example, the consensus of LV-FP, ECFP, Estate1, and Estate2 could raise the F scores as high as 0.99 and 0.99 on ``binding'' and ``uptake'' datasets, respectively.

\subsection{Performance on the extended DAT dataset}

The tests on the benchmark DAT datasets indicate LS-VP and its consensus with 2D fingerprints could yield good predictions. Therefore, these fingerprints were further applied to regression and classification tasks on the extended DAT dataset collected by us from ChEMBL. The training and test sets were also determined by random 85\% : 15\% split. The random splitting, training, and testing were run 35 times to reduce the uncertainty.

\subsubsection{Regression tasks}

\begin{table}
	\centering
	\begin{tabular}{c|c|c|c|c|c}		
		\hline
		
		Model & Metrics & Fingerprint   & Average & Standard deviation & Best   \\ 	
		\hline
		
		\multirow{16}*{\rotatebox{90}{Regression}} & \multirow{8}*{$\rm R^2$}  & {LV-FP}  & 0.66 & 0.03 & 0.73  \\
		\cline{3-6}
		&& ECFP & 0.67 & 0.03  & 0.73 \\
		\cline{3-6}
		&&Estate1&  0.56 & 0.04 & 0.64  \\
		\cline{3-6}
		&&Estate2 &  0.60 & 0.03 & 0.67  \\
		\cline{3-6}
		&&consensus of  &  \multirow{2}*{0.70} & \multirow{2}*{0.03} & \multirow{2}*{0.76}  \\
		&&LV-FP + ECFP  &   &  &   \\
		\cline{3-6}
		&&consensus of &  \multirow{2}*{0.69} & \multirow{2}*{0.03} & \multirow{2}*{0.75}  \\
		&&all 4 FPs  &   &  &   \\
		\cline{2-6}
		
		& \multirow{8}*{$\rm RMSE$}  & {LV-FP}  & 0.87 & 0.04 & 0.80  \\
		\cline{3-6}
		&& ECFP & 0.86 & 0.04  & 0.79 \\
		\cline{3-6}
		&&Estate1&  1.01 & 0.04 & 0.91  \\
		\cline{3-6}
		&&Estate2 &  0.94 & 0.04 & 0.86  \\
		\cline{3-6}
		&&consensus of &  \multirow{2}*{0.82} & \multirow{2}*{0.04} & \multirow{2}*{0.75}  \\
		&&LV-FP + ECFP  &   &  &   \\
		\cline{3-6}
		&&consensus of  &  \multirow{2}*{0.83} & \multirow{2}*{0.04} & \multirow{2}*{0.76}  \\
		&&all 4 FPs  &   &  &   \\
		\cline{2-6}		
		
		\hline
		
	\end{tabular}
	\caption{The regression performance on the extended DAT dataset. The unit of RMSE is kcal/mol.}
	\label{tab:dat-extended-regress}
\end{table}

The whole extended DAT dataset is considered in our regression test. The datset size  is 2878. Our regression performance on this dataset is shown in Table \ref{tab:dat-extended-regress}. 

Among LV-FP, ECFP, Estate1, and Estate2, LV-FP and ECFP are the best with $\rm R^2$s 0.66 and 0.67, respectively. The consensus of LV-FP and ECFP promotes $\rm R^2$ considerably to 0.70. In terms of $\rm RMSE$, the $\rm RMSE$ from LV-FP and ECFP are 0.87 and 0.86. Their consensus reduces $\rm RMSE$ to 0.82.

Notably, even the performances of Estate1 and Estate2 are somewhat poor, but they could play an enhancement role in consensus. For example, the $\rm R^2$ and $\rm RMSE$ by the consensus of four fingerprints are 0.69 and 0.83, respectively, which are still much better than solely from LV-FP or ECFP.

\subsubsection{Classification tasks}
\begin{table}
	\centering
	\begin{tabular}{c|c|c|c|c|c}		
		\hline
		
		Model & Metrics & Fingerprint   & Average & Standard deviation & Best   \\ 	
		\hline
		
		\multirow{16}*{\rotatebox{90}{Classification}} & \multirow{8}*{Accuracy}  & {LV-FP}  & 0.93 & 0.02 & 0.96  \\
		\cline{3-6}
		&& ECFP & 0.93 & 0.02  & 0.97 \\
		\cline{3-6}
		&&Estate1&  0.92 & 0.02 & 0.96  \\
		\cline{3-6}
		&&Estate2 &  0.93 & 0.02 & 0.97  \\
		\cline{3-6}
		&&consensus of &  \multirow{2}*{0.93} & \multirow{2}*{0.02} & \multirow{2}*{0.97}  \\
		&&LV-FP + ECFP  &   &  &   \\
		\cline{3-6}
		&&consensus  &  \multirow{2}*{0.93} & \multirow{2}*{0.02} & \multirow{2}*{0.97}  \\
		&&TOP 4  &   &  &   \\
		\cline{2-6}
		
		& \multirow{8}*{F score}  & {LV-FP}  & 0.98 & 0.01 & 0.99  \\
		\cline{3-6}
		&& ECFP & 0.98 & 0.01  & 0.99 \\
		\cline{3-6}
		&&Estate1&  0.98 & 0.01 & 0.99  \\
		\cline{3-6}
		&&Estate2 &  0.98 & 0.01 & 0.99  \\
		\cline{3-6}
		&&consensus of &  \multirow{2}*{0.98} & \multirow{2}*{0.01} & \multirow{2}*{0.99}  \\
		&&LV-FP + ECFP  &   &  &   \\
		\cline{3-6}
		&&consensus of &  \multirow{2}*{0.98} & \multirow{2}*{0.01} & \multirow{2}*{0.99}  \\
		&&all 4 FPs  &   &  &   \\
		\cline{2-6}		
		
		\hline
		
	\end{tabular}
	\caption{The classification performance on the extended DAT dataset.}
	\label{tab:dat-extended-class}
\end{table}

In the classification test, the compounds with $\rm pK_{i}$ or $\rm pIC_{50}$ between 5 and 6 are removed, and thus 2073 compounds are left. As shown in Table \ref{tab:dat-extended-class}, Since all the fingerprints have very similar performances, consensus does not reflect a noticeable improvement.

\subsection{Predictive power on  D$_2$R inhibitor datasets}

\subsubsection{Classification tasks on the benchmark D$_2$R inhibitor dataset}

For classification tasks, we selected the benchmark D$_2$R inhibitor dataset from Ref. \cite{warszycki2021pharm} and compare our predictions with theirs. For the purpose of comparison, we also performed 10-fold cross-validation and used the literature criteria (MCC and accuracy). Our results are given in Table \ref{tab:d2-class}.

\begin{table}
	\centering
	\begin{tabular}{c|c|c|c|c|c}		
		\hline
		
		Model & Metrics & Fingerprint   & Average & Standard deviation & Best   \\ 	
		\hline
		
		\multirow{18}*{\rotatebox{90}{Classification}} & \multirow{9}*{MCC}  & {LV-FP}  & 0.78 & 0.03 & 0.82  \\
		\cline{3-6}
		&& ECFP & 0.78 & 0.03  & 0.83 \\
		\cline{3-6}
		&&Estate1&  0.70 & 0.01 & 0.72  \\
		\cline{3-6}
		&&Estate2 &  0.74 & 0.04 & 0.80  \\
		\cline{3-6}
		&&consensus of  &  \multirow{2}*{0.80} & \multirow{2}*{0.04} & \multirow{2}*{0.84}  \\
		&&LV-FP + ECFP  &   &  &   \\
		\cline{3-6}
		&&consensus of &  \multirow{2}*{0.76} & \multirow{2}*{0.05} & \multirow{2}*{0.83}  \\
		&&all 4 FPs  &   &  &   \\
		\cline{3-6}
		&&Ref. \cite{warszycki2021pharm}&  0.77 & -- & --  \\
		\cline{2-6}

		& \multirow{9}*{Accuracy}  & {LV-FP}  & 0.90 & 0.01 & 0.92  \\
		\cline{3-6}
		&& ECFP & 0.90 & 0.01  & 0.92 \\
		\cline{3-6}
		&&Estate1&  0.87 & 0.01 & 0.98  \\
		\cline{3-6}
		&&Estate2 &  0.88 & 0.02 & 0.91  \\
		\cline{3-6}
		&&consensus of &  \multirow{2}*{0.91} & \multirow{2}*{0.02} & \multirow{2}*{0.92}  \\
		&&LV-FP + ECFP  &   &  &   \\
		\cline{3-6}
		&&consensus of &  \multirow{2}*{0.88} & \multirow{2}*{0.03} & \multirow{2}*{0.92}  \\
		&&all 4 FPs  &   &  &   \\
		\cline{3-6}	
		&&Ref. \cite{warszycki2021pharm}&  0.89 & -- & --  \\
		\cline{2-6}	
		
		\hline
		
	\end{tabular}
	\caption{The classification performance on the D$_2$R dataset.}
	\label{tab:d2-class}
\end{table}

Our best results are obtained from LV-FP and ECFP, with MCCs of 0.78 and 0.78, respectively. These two results beat the benchmark result (MCC=0.77) from Ref. \cite{warszycki2021pharm}. Our consensus of LV-FP and ECFP promote MCC further to 0.80, which is the best result on this benchmark D$_2$R dataset. The performances of Estate1 and Estate2 are some worse, with MCCs of 0.70 and 0.74, respectively. 

The ranking of accuracy is the same. LV-FP and ECFP yield accuracy as high as 0.90 and 0.90, respectively, which is the same as accuracy=0.90 provided by Ref. \cite{warszycki2021pharm}. The accuracy of their consensus reaches 0.91. The accuracies of Estate1 and Estate2 are 0.87 and 0.88, respectively.

\subsubsection{Regression tasks on the extended D$_2$R inhibitor dataset}

We collected the dataset for  regression tasks from ChEMBL. Our D$_2$R dataset for regression tasks has 6943 compounds. Our performance on this dataset is depicted in Table \ref{tab:d2-regress}.

Similar to the results from other datasets,  ECFP and LV-FP have the best performance  with $\rm R^2$s 0.64 and 0.68, respectively. The consensus of LV-FP and ECFP further lifts the $\rm R^2$ to 0.69. The results from Estate1 and Estate2 are somewhat poor with $\rm R^2$s 0.55 and 0.60. However, their consensus including Estate1 and Estate2 still leads to a promising $\rm R^2$ of 0.67.

The same trend is found for RMSEs. The RMSEs from ECFP and LV-FP are the lowest: 0.79 and 0.87. Their consensus also has a RMSE of 0.79. The RMSEs from Estate1 and Estate2 are 0.94 and 0.89, respectively. 

\begin{table}
	\centering
	\begin{tabular}{c|c|c|c|c|c}		
		\hline
		
		Model & Metrics & Fingerprint   & Average & Standard deviation & Best   \\ 	
		\hline
		
		\multirow{16}*{\rotatebox{90}{Regression}} & \multirow{8}*{$\rm R^2$}  & {LV-FP}  & 0.64 & 0.02 & 0.66  \\
		\cline{3-6}
		&& ECFP & 0.68 & 0.02  & 0.70 \\
		\cline{3-6}
		&&Estate1&  0.55 & 0.03 & 0.59  \\
		\cline{3-6}
		&&Estate2 &  0.60 & 0.03 & 0.63  \\
		\cline{3-6}
		&&consensus of  &  \multirow{2}*{0.69} & \multirow{2}*{0.02} & \multirow{2}*{0.71}  \\
		&&LV-FP + ECFP  &   &  &   \\
		\cline{3-6}
		&&consensus of &  \multirow{2}*{0.67} & \multirow{2}*{0.02} & \multirow{2}*{0.71}  \\
		&&all 4 FPs  &   &  &   \\
		\cline{2-6}
		
		& \multirow{8}*{$\rm RMSE$}  & {LV-FP}  & 0.87 & 0.03 & 0.83  \\
		\cline{3-6}
		&& ECFP & 0.79 & 0.04  & 0.75 \\
		\cline{3-6}
		&&Estate1&  0.94 & 0.04 & 0.90  \\
		\cline{3-6}
		&&Estate2 &  0.89 & 0.03 & 0.85  \\
		\cline{3-6}
		&&consensus of &  \multirow{2}*{0.79} & \multirow{2}*{0.03} & \multirow{2}*{0.73}  \\
		&&LV-FP + ECFP  &   &  &   \\
		\cline{3-6}
		&&consensus of &  \multirow{2}*{0.82} & \multirow{2}*{0.03} & \multirow{2}*{0.78}  \\
		&&all 4 FPs  &   &  &   \\
		\cline{2-6}		
		
		\hline

	\end{tabular}
	\caption{The regression performance on the D$_2$R dataset. The unit of RMSE is kcal/mol.}
	\label{tab:d2-regress}
\end{table}

\subsection{Our predictive power on our D$_3$R inhibitor dataset}

We compiled our D$_3$R inhibitor dataset from ChEMBL and tested the predictive power of our models for both regression and classification tasks of this dataset. Here we performed 10-fold cross-validation.

\subsubsection{Regression tasks}

The D$_3$R dataset for regression tasks contains 4685 compounds. Table \ref{tab:d3-regress} shows our results.

Similar to the performance on the DAT datasets, Among LV-FP, ECFP, Estate1, and Estate2, ECFP and LV-FP are the best two fingerprints with $\rm R^2$s 0.70 and 0.66. The consensus of LV-FP and ECFP further lifts the $\rm R^2$ to 0.71. The results from Estate1 and Estate2 are somewhat poor with $\rm R^2$s 0.59 and 0.64. However, their consensus with ECFP and LV-FP is still as good as the best single fingerprint ECFP, which suggests considering them in the consensus is still helpful for improving the results. 

The RMSEs tell the same trend. The RMSEs from ECFP and LV-FP are the lowest: 0.86 and 0.94. The RMSEs from Estate1 and Estate2 are 1.02 and 0.95, respectively. But the consensus including Estate1 and Estate2 still achieves a promising RMSE of 0.87. 

\begin{table}
	\centering 
	\begin{tabular}{c|c|c|c|c|c}		
		\hline
		
		Model & Metrics & Fingerprint   & Average & Standard deviation & Best   \\ 	
		\hline
		
		\multirow{16}*{\rotatebox{90}{Regression}} & \multirow{8}*{$\rm R^2$}  & {LV-FP}  & 0.66 & 0.02 & 0.70  \\
		\cline{3-6}
		&& ECFP & 0.70 & 0.02  & 0.73 \\
		\cline{3-6}
		&&Estate1&  0.59 & 0.03 & 0.64  \\
		\cline{3-6}
		&&Estate2 &  0.64 & 0.02 & 0.68  \\
		\cline{3-6}
		&&consensus of &  \multirow{2}*{0.71} & \multirow{2}*{0.02} & \multirow{2}*{0.74}  \\
		&&LV-FP + ECFP  &   &  &   \\
		\cline{3-6}
		&&consensus of &  \multirow{2}*{0.70} & \multirow{2}*{0.02} & \multirow{2}*{0.74}  \\
		&& all 4 FPs  &   &  &   \\
		\cline{2-6}
		
		& \multirow{8}*{$\rm RMSE$}  & {LV-FP}  & 0.94 & 0.03 & 0.89  \\
		\cline{3-6}
		&& ECFP & 0.86 & 0.03  & 0.83 \\
		\cline{3-6}
		&&Estate1&  1.02 & 0.04 & 0.95  \\
		\cline{3-6}
		&&Estate2 &  0.95 & 0.03 & 0.91  \\
		\cline{3-6}
		&&consensus of &  \multirow{2}*{0.86} & \multirow{2}*{0.03} & \multirow{2}*{0.83}  \\
		&&LV-FP + ECFP  &   &  &   \\
		\cline{3-6}
		&&consensus of &  \multirow{2}*{0.87} & \multirow{2}*{0.02} & \multirow{2}*{0.61}  \\
		&&all 4 FPs  &   &  &   \\
		\cline{2-6}		
		
		\hline

	\end{tabular}
	\caption{The regression performance on the D$_3$R dataset. The unit of RMSE is kcal/mol.}
	\label{tab:d3-regress}
\end{table}

\subsubsection{Classification tasks}

We applied the same procedure to  D2-inhibitors as reported in the literature \cite{warszycki2021pharm}, which eliminates the compounds with $\rm pK_{i}$ or $\rm pIC_{50}$ between 6 and 7 to define an obvious boundary between actives and inactives. As a result, the D$_3$R classification dataset is downsized, which contains 3695 compounds. Our performance  is given in Table \ref{tab:d3-class}.

\begin{table}
	\centering
	\begin{tabular}{c|c|c|c|c|c}		
		\hline
		
		Model & Metrics & Fingerprint   & Average & Standard deviation & Best   \\ 	
		\hline
		
		\multirow{16}*{\rotatebox{90}{Classification}} & \multirow{8}*{Accuracy}  & {LV-FP}  & 0.92 & 0.01 & 0.94  \\
		\cline{3-6}
		&& ECFP & 0.93 & 0.02  & 0.95 \\
		\cline{3-6}
		&&Estate1&  0.91 & 0.02 & 0.93  \\
		\cline{3-6}
		&&Estate2 &  0.92 & 0.01 & 0.94  \\
		\cline{3-6}
		&&consensus of  &  \multirow{2}*{0.94} & \multirow{2}*{0.01} & \multirow{2}*{0.95}  \\
		&&LV-FP + ECFP  &   &  &   \\
		\cline{3-6}
		&&consensus of  &  \multirow{2}*{0.93} & \multirow{2}*{0.01} & \multirow{2}*{0.94}  \\
		&& all 4 FPs  &   &  &   \\
		\cline{2-6}
		
		& \multirow{8}*{F score}  & {LV-FP}  & 0.95 & 0.01 & 0.96  \\
		\cline{3-6}
		&& ECFP & 0.96 & 0.01  & 0.97 \\
		\cline{3-6}
		&&Estate1&  0.95 & 0.01 & 0.96  \\
		\cline{3-6}
		&&Estate2 &  0.95 & 0.01 & 0.96  \\
		\cline{3-6}
		&&consensus of &  \multirow{2}*{0.96} & \multirow{2}*{0.01} & \multirow{2}*{0.97}  \\
		&&LV-FP + ECFP  &   &  &   \\
		\cline{3-6}
		&&consensus of &  \multirow{2}*{0.96} & \multirow{2}*{0.01} & \multirow{2}*{0.97}  \\
		&& all 4 FPs  &   &  &   \\
		\cline{2-6}		
		
		\hline
		
	\end{tabular}
	\caption{The classification performance on the D$_3$R dataset.}
	\label{tab:d3-class}
\end{table}

As expected, LV-FP and ECFP still outperform other 2D fingerprints with accuracy values of 0.92 and 0.93. The consensus of LV-FP and ECFP provides an enhancement to 0.94 (consensus of top 2 in Table \ref{tab:d3-class}). Estate1 and Estate2 obtain an accuracy of 0.91 and 0.92, respectively. The consensus involving all the four fingerprints could reach an accuracy value as high as 0.93.

In terms of F scores, all the four 2D fingerprints LV-FP, ECFP, Estate1, and Estate2 lead to very similar values: 0.95, 0.96, 0.95, and 0.95, respectively. The consensuses from LV-FP and ECFP or all the four fingerprints also have a F score of 0.96.

\subsection{Our performance on the benchmark hERG dataset from previous work}

The benchmark hERG dataset is provided in Lee et al.'s recent work \cite{fant2019toward}. We follow the splitting between training and test given by Lee et al. The training set and test set are generated by a random split with a ratio of  85\% to 15\%. We repeated splitting, training, and testing 35 times to reduce uncertainty.

We tested these datasets on both regression and classification tasks. Since in the classification tasks, the data points with $\rm pK_{i}$ or $\rm pIC_{50}$ between 5 and 6 were erased to define a clear boundary between actives and inactives. The dataset for classification are smaller than that for regression.

\subsubsection{Regression tasks}

As shown in Table \ref{tab:herg-benckmark-regress}, in terms of $\rm R^2$, between LV-FP, ECFP, Estate1, and Estate2, LV-FP outperforms other three traditional 2D fingerprints. With $\rm R^2$=0.66, the performance of LV-FP is as good as that reported by the benchmark work \cite{fant2019toward}. The consensus containing and LV-FP and other fingerprints could further improve the results. Notably, although the performance of Estate1 or Estate2 alone is limited, the consensus containing them is better than any single fingerprint.

In terms of RMSE, our consensus results (0.76) are the best among all ours and that from Ref. \cite{fant2019toward}, and provide remarkable enhancement from any single fingerprint of ours. The best single fingerprint leads to an RMSE of 0.80.

\begin{table}
	\centering
	\begin{tabular}{c|c|c|c|c|c}		
		\hline
		
		Model & Metrics & Fingerprint   & Average & Standard deviation & Best   \\ 	
		\hline
		
		\multirow{18}*{\rotatebox{90}{Regression}} & \multirow{9}*{$\rm R^2$}  & {LV-FP}  & 0.66 & 0.05 & 0.74  \\
		\cline{3-6}
		&& ECFP & 0.63 & 0.04  & 0.74 \\
		\cline{3-6}
		&&Estate1&  0.58 & 0.05 & 0.65  \\
		\cline{3-6}
		&&Estate2 &  0.58 & 0.04 & 0.63  \\
		\cline{3-6}
		&&consensus of &  \multirow{2}*{0.68} & \multirow{2}*{0.03} & \multirow{2}*{0.76}  \\
		&&LV-FP + ECFP  &   &  &   \\
		\cline{3-6}
		&&consensus of &  \multirow{2}*{0.67} & \multirow{2}*{0.03} & \multirow{2}*{0.74}  \\
		&&all 4 FPs  &   &  &   \\
		\cline{3-6}
		&&Ref. \cite{fant2019toward}&  0.66 & 0.04 & 0.76  \\
		\cline{2-6}
		
		& \multirow{9}*{$\rm RMSE$}  & {LV-FP}  & 0.80 & 0.05 & 0.74  \\
		\cline{3-6}
		&& ECFP & 0.80 & 0.05  & 0.70 \\
		\cline{3-6}
		&&Estate1&  0.87 & 0.04 & 0.76  \\
		\cline{3-6}
		&&Estate2 &  0.87 & 0.04 & 0.78  \\
		\cline{3-6}
		&&consensus of &  \multirow{2}*{0.76} & \multirow{2}*{0.04} & \multirow{2}*{0.67}  \\
		&&LV-FP + ECFP  &   &  &   \\
		\cline{3-6}
		&&consensus of &  \multirow{2}*{0.76} & \multirow{2}*{0.04} & \multirow{2}*{0.67}  \\
		&&all 4 FPs  &   &  &   \\
		\cline{3-6}
		&&Ref. \cite{fant2019toward} &  0.79 & 0.04 & --  \\
		\cline{2-6}				
		\hline		
		
	\end{tabular}
	\caption{The regression results on the benchmark hERG datasets. The unit of RMSE is kcal/mol.}
	\label{tab:herg-benckmark-regress}
\end{table}

\subsubsection{Classification tasks}

\begin{table}
	\centering
	\begin{tabular}{c|c|c|c|c|c}		
		\hline
		
		Model & Metrics & Fingerprint   & Average & Standard deviation & Best   \\ 	
		\hline
		
		\multirow{18}*{\rotatebox{90}{Classification}} & \multirow{9}*{Accuracy}  & {LV-FP}  & 0.89 & 0.03 & 0.95  \\
		\cline{3-6}
		&& ECFP & 0.88 & 0.02  & 0.92 \\
		\cline{3-6}
		&&Estate1&  0.87 & 0.02 & 0.91  \\
		\cline{3-6}
		&&Estate2 &  0.87 & 0.03 & 0.92  \\
		\cline{3-6}
		&&consensus of &  \multirow{2}*{0.88} & \multirow{2}*{0.02} & \multirow{2}*{0.92}  \\
		&&LV-FP + ECFP  &   &  &   \\
		\cline{3-6}
		&&consensus of &  \multirow{2}*{0.87} & \multirow{2}*{0.02} & \multirow{2}*{0.91}  \\
		&&all 4 FPs  &   &  &   \\
		\cline{3-6}
		&&Ref. \cite{fant2019toward}&  0.89 & 0.03 & --  \\
		\cline{2-6}
		
		& \multirow{9}*{F score}  & {LV-FP}  & 0.88 & 0.03 & 0.95  \\
		\cline{3-6}
		&& ECFP & 0.88 & 0.02  & 0.92 \\
		\cline{3-6}
		&&Estate1&  0.87 & 0.02 & 0.91  \\
		\cline{3-6}
		&&Estate2 &  0.87 & 0.03 & 0.92  \\
		\cline{3-6}
		&&consensus of &  \multirow{2}*{0.88} & \multirow{2}*{0.02} & \multirow{2}*{0.92}  \\
		&&LV-FP + ECFP  &   &  &   \\
		\cline{3-6}
		&&consensus of &  \multirow{2}*{0.87} & \multirow{2}*{0.02} & \multirow{2}*{0.91}  \\
		&& all 4 FPs  &   &  &   \\
		\cline{3-6}
		&&Ref. \cite{fant2019toward}&  0.88 & 0.03 & --  \\
		\cline{2-6}		
		
		\hline

	\end{tabular}
	\caption{The classification results on the benchmark hERG datasets.}
	\label{tab:herg-benckmark-class}
\end{table}

For classification tasks, since the data points with $\rm pK_{i}$ or $\rm pIC_{50}$ between 5 and 6 were removed, the data number is reduced to 1137 from 2043. Table \ref{tab:herg-benckmark-class} shows our classification results.

As the same as the regression tasks, the results from LV-FP are also some better than our other results and as good as that from Ref. \cite{fant2019toward}. The accuracy and F score from LV-FP are 0.89 and 0.88, respectively.

\subsection{Performance on our extended hERG dataset}

We also developed our extended dataset from ChEMBL.    The training and test sets are generated by a random split with a ratio of  85\% to 15\% as in the literature \cite{fant2019toward}. We repeated splitting, training, and testing 35 times to reduce   uncertainty.

\subsubsection{Regression tasks}

As shown in Table \ref{tab:extend-herg-regress}, in terms of $\rm R^2$, the consensus containing and LV-FP and ECFP could yield the best results as high as 0.64. Notably, although the performance of Estate1 or Estate2 alone is limited, the consensus containing them can reach 0.63.

In terms of RMSE, our consensus results are the best (0.74). The best single fingerprint could lead to an RMSE of 0.78.

\begin{table}
	\centering
	\begin{tabular}{c|c|c|c|c|c}		
		\hline
		
		Model & Metrics & Fingerprint   & Average & Standard deviation & Best   \\ 	
		\hline
		
		\multirow{18}*{\rotatebox{90}{Regression}} & \multirow{9}*{$\rm R^2$}  & {LV-FP}  & 0.59 & 0.02 & 0.63  \\
		\cline{3-6}
		&& ECFP & 0.60 & 0.02  & 0.65 \\
		\cline{3-6}
		&&Estate1&  0.51 & 0.03 & 0.56  \\
		\cline{3-6}
		&&Estate2 &  0.55 & 0.02 & 0.59  \\
		\cline{3-6}
		&&consensus of  &  \multirow{2}*{0.64} & \multirow{2}*{0.02} & \multirow{2}*{0.68}  \\
		&&LV-FP + ECFP  &   &  &   \\
		\cline{3-6}
		&&consensus of  &  \multirow{2}*{0.63} & \multirow{2}*{0.02} & \multirow{2}*{0.66}  \\
		&&all 4 FPs  &   &  &   \\
		\cline{2-6}
		
		& \multirow{9}*{$\rm RMSE$}  & {LV-FP}  & 0.78 & 0.03 & 0.74  \\
		\cline{3-6}
		&& ECFP & 0.76 & 0.03  & 0.72 \\
		\cline{3-6}
		&&Estate1&  0.86 & 0.03 & 0.79  \\
		\cline{3-6}
		&&Estate2 &  0.82 & 0.03 & 0.76  \\
		\cline{3-6}
		&&consensus of &  \multirow{2}*{0.74} & \multirow{2}*{0.03} & \multirow{2}*{0.70}  \\
		&&LV-FP + ECFP  &   &  &   \\
		\cline{3-6}
		&&consensus of  &  \multirow{2}*{0.74} & \multirow{2}*{0.03} & \multirow{2}*{0.70}  \\
		&& all 4 FPs  &   &  &   \\
		\cline{2-6}				
		\hline		
		
	\end{tabular}
	\caption{The regression results on the extended hERG dataset.}
	\label{tab:extend-herg-regress}
\end{table}

\subsubsection{Classification tasks}

For classification tasks, since the data points with $\rm pK_{i}$ or $\rm pIC_{50}$ between 5 and 6 were removed, the data number is reduced to 3117 from 6298. Table \ref{tab:extend-herg-class} shows the classification results.

As the same as the regression tasks, LV-FP and ECFP are also better than others. Their consensus can further improve the accuracy and F score to 0.88 and 0.91, respectively. 

\begin{table}
	\centering
	\begin{tabular}{c|c|c|c|c|c}		
		\hline
		
		Model & Metrics & Fingerprint   & Average & Standard deviation & Best   \\ 	
		\hline
		
		\multirow{18}*{\rotatebox{90}{Classification}} & \multirow{9}*{Accuracy}  & {LV-FP}  & 0.87 & 0.02 & 0.90  \\
		\cline{3-6}
		&& ECFP & 0.87 & 0.02  & 0.90 \\
		\cline{3-6}
		&&Estate1&  0.87 & 0.02 & 0.90  \\
		\cline{3-6}
		&&Estate2 &  0.87 & 0.03 & 0.92  \\
		\cline{3-6}
		&&consensus of  &  \multirow{2}*{0.88} & \multirow{2}*{0.02} & \multirow{2}*{0.91}  \\
		&&LV-FP + ECFP  &   &  &   \\
		\cline{3-6}
		&&consensus of &  \multirow{2}*{0.87} & \multirow{2}*{0.02} & \multirow{2}*{0.90}  \\
		&&all 4 FPs  &   &  &   \\
		\cline{2-6}
		
		& \multirow{9}*{F score}  & {LV-FP}  & 0.90 & 0.01 & 0.93  \\
		\cline{3-6}
		&& ECFP & 0.90 & 0.01  & 0.93 \\
		\cline{3-6}
		&&Estate1&  0.90 & 0.01 & 0.93  \\
		\cline{3-6}
		&&Estate2 &  0.90 & 0.01 & 0.92  \\
		\cline{3-6}
		&&consensus of  &  \multirow{2}*{0.91} & \multirow{2}*{0.01} & \multirow{2}*{0.93}  \\
		&&LV-FP + ECFP  &   &  &   \\
		\cline{3-6}
		&&consensus of &  \multirow{2}*{0.90} & \multirow{2}*{0.02} & \multirow{2}*{0.92}  \\
		&&all 4 FPs  &   &  &   \\
		\cline{2-6}		
		
		\hline

	\end{tabular}
	\caption{The classification results on the extended hERG datasets.}
	\label{tab:extend-herg-class}
\end{table}


\clearpage

\bibliographystyle{unsrt}
\bibliography{refs}